\documentclass[twocolumn,pre,twoside,superscriptaddress,floatfix]{revtex4}

\usepackage{amsmath}
\usepackage{amssymb}
\usepackage{graphicx}
\usepackage{dcolumn}
\usepackage{bm}
\usepackage{color}

\def\dps{\displaystyle}
\def\m#1{\mathrm{#1}}
\def\Eq#1{(\ref{eq:#1})}
\def\d{\mathrm{d}}

\def\epsilon{\varepsilon}
\def\theta{\vartheta}
\def\rho{\varrho}
\def\Int#1#2#3{\int_{#1}\!\mathrm{d}^{#2}{#3}\;}

\def\vec#1{\mathbf{#1}}
\def\Re{\mathop\mathrm{Re}}
\def\Im{\mathop\mathrm{Im}}
\def\sign{\mathop\mathrm{sign}}
\def\artanh{\mathop\mathrm{artanh}}
\def\arsinh{\mathop\mathrm{arsinh}}

\begin{document}


\title{Local theory for ions in binary liquid mixtures}

\author{Markus Bier}
\email{bier@is.mpg.de}
\affiliation
{
   Max-Planck-Institut f\"ur Intelligente Systeme, 
   Heisenbergstr.\ 3,
   70569 Stuttgart,
   Germany, 
   and
   Institut f\"ur Theoretische und Angewandte Physik,
   Universit\"at Stuttgart,
   Pfaffenwaldring 57,
   70569 Stuttgart,
   Germany
}
\author{Andrea Gambassi}
\affiliation
{
   SISSA --- International School for Advanced Studies and INFN,
   via Bonomea 265,
   34136 Trieste,
   Italy
}
\author{S.\ Dietrich}
\affiliation
{
   Max-Planck-Institut f\"ur Intelligente Systeme, 
   Heisenbergstr.\ 3,
   70569 Stuttgart,
   Germany, 
   and
   Institut f\"ur Theoretische und Angewandte Physik,
   Universit\"at Stuttgart,
   Pfaffenwaldring 57,
   70569 Stuttgart,
   Germany
}

\date{6 July, 2012}

\begin{abstract}
The influence of ions on the bulk phase behavior of binary liquid mixtures acting as their solvents and on the
corresponding interfacial structures close to a planar wall is investigated by means of density functional theory
based on local descriptions of the effective interactions between ions and their solvents.
The bilinear coupling approximation (BCA), which has been used in numerous previous related investigations, is compared 
with a novel local density approximation (LDA) for the ion-solvent interactions.
It turns out that within BCA the bulk phase diagrams, the two-point correlation functions, and critical adsorption exhibit
 qualitative features which are not compatible with the available experimental data.
These discrepancies do not occur within the proposed LDA.
Further experimental investigations are suggested which assess the reliability of the proposed LDA. 
This approach allows one to obtain a consistent and rather general understanding of the effects of ions on solvent 
properties.
From our analysis we infer in particular that there can be an experimentally detectable influence of ions on binary liquid 
mixtures due to steric effects but not due to charge effects.
\end{abstract}

\maketitle


\section{Introduction}

Ion-solvent mixtures play a central role for various important soft matter systems such as colloidal suspensions, 
polymer solutions, biochemical reactions, and electrochemical cells.
For most of these systems the presence of an appropriate amount of ions is crucial for their functioning.
Therefore the general question arises of how and to what extent ions alter the properties of a solvent.
More than a century ago the considerations by Arrhenius concerning the colligative properties of ionic solutions 
\cite{Arrhenius1887} seemed to imply that the interactions of ions with their fluid environment have only a mild effect
in the sense that adding salt merely modifies the composition of the ion-solvent mixture.
A few decades later, after more accurate measurements had been carried out, this simple picture was questioned.
Bjerrum, Debye, and H\"{u}ckel pointed out that electrostatically induced ion-ion correlations are expected to 
influence the ion distribution and the equation of state of electrolyte solutions \cite{Debye1923,McQuarrie2000}.
Since then, the Debye-H\"{u}ckel theory has been widely used, e.g., in plasma physics and as an ingredient
of the DLVO theory (named after Derjaguin, Landau, Verwey, and Overbeek) of colloidal suspensions \cite{Russel1989}.
Within Debye-H\"{u}ckel theory the solvent is considered to be a uniform dielectric continuum which influences the
Coulomb interaction between the ions via a certain permittivity but which is otherwise inert.
Recently the issue of the mutual influence of ions and the solvent has been taken up again, 
studying the solubility of ions and the double-layer structure in a near-critical solvent 
\cite{Onuki2004,Onuki2006,Okamoto2010,Okamoto2011,Ciach2010}, possible salt-induced changes of the solvent structure 
\cite{Nabutovskii1980,Nabutovskii1985,Nabutovskii1986,Sadakane2006,Sadakane2007a,Sadakane2007b,Sadakane2009,Sadakane2011}, 
and effects of the inhomogeneities of the permittivity close to interfaces 
\cite{Tsori2007,BenYaakov2009,Oleksy2009,Samin2011a}.

These investigations require a model for the solvent at least on the mesoscopic scale as well as a description of the 
ion-solvent interaction.
To this end one can split the pair potential between the species into the long-ranged electrostatic monopole-monopole 
contribution and into the remaining contributions of shorter range, which we shall refer to as chemical contributions.
In the vast majority of the theoretical studies of dilute electrolyte solutions, ions are described as point-like 
particles whose chemical contributions to the interactions with the solvent are modeled locally within the so-called 
bilinear coupling approximation (BCA).
This amounts to a local density approximation for the chemical contribution to the excess free energy which is bilinear
in the particle number densities 
\cite{Onuki2004,Onuki2006,Okamoto2010,Okamoto2011,Nabutovskii1980,Nabutovskii1985,Nabutovskii1986,Tsori2007,%
BenYaakov2009,Samin2011a}.
Within the approaches of Refs.~\cite{Ciach2010} and \cite{Oleksy2009} the ion size is accounted for by means of hard-core
exclusion and solvation is modeled by non-local interactions within the so-called random-phase approximation (RPA) for
density functional theory (DFT).
In fact, the BCA can be considered as the local version of the RPA, which is expected to be reliable only for interaction
energies small compared with the thermal energy \cite{Hansen1986}.
However, the ion-solvent interaction is typically of the order of some tens of the thermal energy 
\cite{Marcus1983,Inerowicz1994}.
Therefore the application of the BCA or of the RPA to ion-solvent mixtures is questionable \cite{CommentOnuki2006}.

In the following we show that for realistic values of the parameters the BCA and the RPA indeed lead to unphysical results.
Since non-local models are notoriously complicated it seems worthwhile to investigate the possibility of local 
descriptions of ion-solvent interactions.
In order to demonstrate that valuable improvements within the class of local models are possible, we propose an 
alternative local density approximation (LDA) the predictions of which are in qualitative agreement with experimental 
results and which does not lead to the artifacts introduced by the BCA.
Without presenting the details which will be expounded below, this LDA has already been applied successfully in a recent 
study of the effective interaction between two planar substrates in contact with a near-critical binary liquid mixture and in 
the presence of salt \cite{Bier2011}.
As far as the phase behavior, the bulk structure, and the asymptotic interfacial structure of the solvent are 
concerned, our results are equivalent to replacing the bare interaction energies used within the BCA by effective
ones which saturate for large values of the bare interaction energies.
Within this LDA, and in contrast to the BCA, more realistic estimates of the magnitude of salt-induced effects
can be obtained, which are expected to be important for interpreting and designing experiments and for considering 
applications.

In the following, after introducing the model and the LDA in Sec.~\ref{Sec:Model} and in Appendix~\ref{AppA}, bulk 
systems are discussed in Sec.~\ref{Sec:Bulk}, where we focus on the phase diagram and on the structure of the correlation
functions. 
Similarities and differences between the proposed LDA and the BCA are highlighted and compared with the available 
experimental data. 
In Sec.~\ref{Sec:Interface} interfacial structures as well as critical adsorption in semi-infinite planar systems are 
discussed.
For critical adsorption important qualitative differences between the LDA and the BCA are revealed, with the predictions
of the former being in agreement with the corresponding qualitative experimental findings, whereas those of the latter are not.
We propose further experimental investigations which are expected to discriminate more sharply between the LDA and the BCA
than the presently available data do.
In Sec.~\ref{Sec:Colloids} an application of the proposed LDA to colloidal interactions in near-critical electrolyte 
solutions is discussed and compared with alternative approaches described in the literature.
Finally, in Sec.~\ref{Sec:Conclusions} we draw our conclusions and provide a summary.


\section{\label{Sec:Model}Model}

\subsection{Definition}

We consider a three-dimensional ($d=3$) container $\widetilde{\mathcal{V}}\subseteq\mathbb{R}^3$
filled with an incompressible binary liquid mixture acting as a solvent for cations ($+$) and anions ($-$).
All solvent particles are assumed to be of equal size with non-vanishing volume $\widetilde{a}^3$ whereas the
ions are considered to be point-like; hence ions do not contribute to the total packing fraction (see also
Appendix \ref{AppA}).
The set of dimensionless positions $\vec{r}=(x,y,z):=\widetilde{\vec{r}}/\widetilde{a}$ for 
$\widetilde{\vec{r}}\in\widetilde{\mathcal{V}}$ is defined as $\mathcal{V}$.
At $\vec{r}\in\mathcal{V}$ the number densities of the solvent components $A$ and $B$ are given by
$\widetilde{\rho}_A(\vec{r})=\phi(\vec{r})\widetilde{a}^{-3}$ and $\widetilde{\rho}_B(\vec{r})=
(1-\phi(\vec{r}))\widetilde{a}^{-3}$, respectively, with $0\leq\phi\leq1$, whereas the number densities 
of the cations and anions are given by $\widetilde{\rho}_+(\vec{r})=\rho_+(\vec{r})\widetilde{a}^{-3}$ and 
$\widetilde{\rho}_-(\vec{r})=\rho_-(\vec{r})\widetilde{a}^{-3}$, respectively.
The walls $\partial\mathcal{V}$ of the container carry a surface charge density
$\sigma(\vec{r})e\widetilde{a}^{-2}$ at $\vec{r}\in\partial\mathcal{V}$, where $e$ is the (positive)
elementary charge.
The influence of the walls onto the solvent due to short-ranged chemical effects is captured by surface fields localized
at the walls.
At $\vec{r}\in \partial\mathcal{V}$, the dimensionless volume fraction $\phi(\vec{r})$ of $A$ particles couples
linearly to surface fields $h(\vec{r})$, where $h>0$ ($<0$) leads to a preferential adsorption of solvent component 
$A$ ($B$).
The equilibrium profiles $\phi$, $\rho_+$, and $\rho_-$ minimize the approximate grand potential
density functional $k_BT\Omega[\phi,\rho_\pm]$,
\begin{eqnarray}
   \Omega[\phi,\rho_\pm] 
   & = & 
   \Int{\mathcal{V}}{3}{r} 
   \bigg\{\omega_\m{sol}(\phi(\vec{r})) + \frac{\chi(T)}{6}(\nabla\phi(\vec{r}))^2 
   \nonumber\\
   && 
   + \sum_{i=\pm}\Big[\omega^{(i)}_\m{ion}(\rho_i(\vec{r})) + 
                      \rho_i(\vec{r})V_i(\phi(\vec{r}))\Big]
   \label{eq:df}\\
   && 
   + \frac{2\pi\ell_B}{\epsilon(\phi(\vec{r}))}\vec{D}(\vec{r},[\rho_\pm])^2\bigg\}
   - \Int{\partial\mathcal{V}}{2}{r} h(\vec{r})\phi(\vec{r}),
   \nonumber
\end{eqnarray}
with $\omega_\m{sol}(\phi)=\phi(\ln\phi-\mu_\phi) + (1-\phi)\ln(1-\phi) + \chi(T)\phi(1-\phi)$ 
and $\omega^{(\pm)}_\m{ion}(\rho_\pm)=\rho_\pm(\ln\rho_\pm - 1 - \mu_\pm)$ as the bulk grand potential
densities of the solvent and of the $\pm$-ions (in the low number density limit), respectively.
Here $k_BT$ is the thermal energy, $\mu_\phi k_BT$ and $\mu_\pm k_BT$ are the chemical potential difference 
$(\mu_A-\mu_B)k_BT$ of the solvent particles and the chemical potentials of the $\pm$-ions, respectively, and
$\ell_B\widetilde{a}=e^2/(4\pi\epsilon_0 k_BT)$ is the Bjerrum length for the vacuum permittivity
$\epsilon_0$.
The temperature-dependent Flory-Huggins parameter $\chi(T)>0$ describes the effective interaction between
solvent particles, where the temperature dependence is usually described by the empirical form
$\dps \chi(T)=\chi_S+\frac{\chi_H}{T}$ with the system specific entropic contribution $\chi_S$ and the enthalpic 
contribution $\chi_H$ \cite{Rubinstein2004}.
For $\chi(T)\geq\chi(T_c)$ phase separation occurs in the pure, salt-free solvent within a certain range of $\phi$ 
whereas for $\chi(T)<\chi(T_c)$ the solvent components $A$ and $B$ are miscible in any proportion.
A positive (negative) enthalpic contribution $\chi_H$ corresponds to an upper (lower) critical demixing point.
The gradient term $\propto(\nabla\phi(z))^2$ with $\nabla = \widetilde{a}\widetilde{\nabla}$ penalizes the spatial
variation of the solvent composition \cite{Cahn1958}.
The ion-solvent interaction is described within a local density approximation (LDA) by the effective 
ion potential $k_BTV_\pm(\phi)$ generated by the solvent (see below).
The relative permittivity $\epsilon(\phi(\vec{r}))$ is assumed to depend locally on the composition
of the solvent $\phi(\vec{r})$ but not on the ion densities $\rho_\pm(\vec{r})$, which is justified 
for small ionic strengths, i.e., $\rho_\pm(\vec{r})\ll1$.
Here the mixing formula $\epsilon(\phi)=\epsilon_A\phi + \epsilon_B(1-\phi)$ introduced by B\"{o}ttcher
\cite{Bottcher1973} is used \cite{Onuki2004,BenYaakov2009,Samin2011a}.
Using SI-units, the electric displacement $\widetilde{\vec{D}}=\vec{D}e\widetilde{a}^{-2}$ in Eq.~\Eq{df}
fulfills Gauss' law $\nabla\cdot\vec{D}(\vec{r},[\rho_\pm])=\rho_+(\vec{r})-\rho_-(\vec{r}),
\vec{r}\in\mathcal{V}$, with fixed surface charges 
$\vec{n}(\vec{r})\cdot\vec{D}(\vec{r},[\rho_\pm])=\sigma(\vec{r}), 
\vec{r}\in\partial\mathcal{V}$, where $\vec{n}$ is the unit vector perpendicular to 
$\partial\mathcal{V}$ pointing towards the exterior of $\mathcal{V}$ (see Ref.~\cite{Russel1989}).
Note that $\vec{D}(\vec{r},[\rho_\pm])$ is generated by the $\pm$-ions and the given surface
charges $\sigma$; it does not depend explicitly on $\phi$.
Within the present model, besides being confined, ions interact with the walls only electrostatically.

\begin{figure*}[!t]
   \includegraphics[width=16cm]{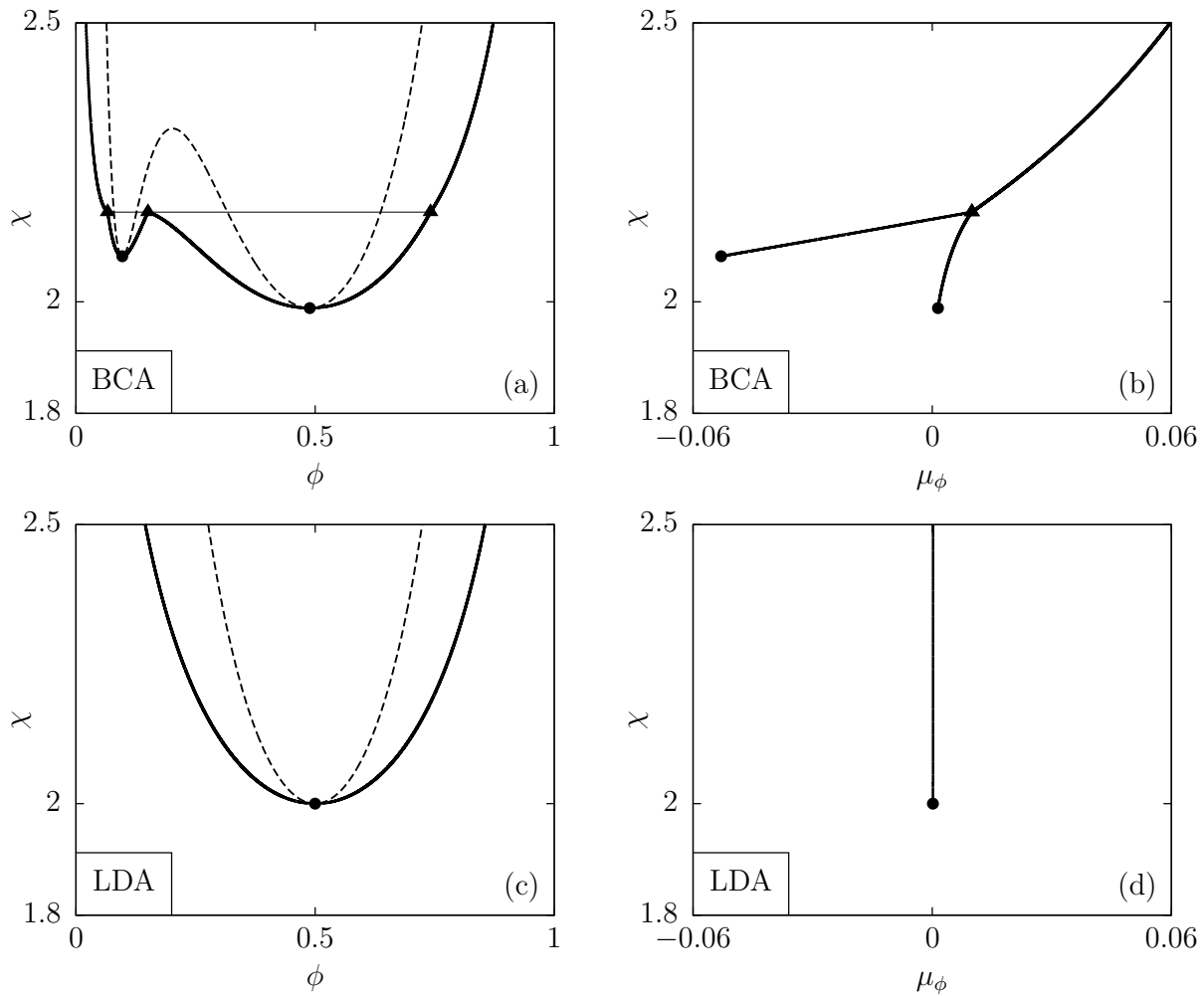}
   \caption{Bulk phase diagrams of a binary liquid mixture with added salt of constant chemical potential (per $k_BT$)
            $\mu_I=\mu_++\mu_-$ within the bilinear coupling approximation (BCA) 
            $V^{(BCA)}_\pm(\phi)=f_\pm\phi$ [(a) and (b)] and within the local density approximation (LDA, see Eq.~\Eq{V}) 
            $V_\pm(\phi)=-\ln(1-\phi(1-\exp(-f_\pm)))$ [(c) and (d)] in terms of the Flory-Huggins 
            parameter $\chi$ and the composition $\phi$ [(a) and (c)] or the chemical potential (per $k_BT$) 
            $\mu_\phi=\mu_A-\mu_B$ conjugate to the composition $\phi$ of the binary solvent [(b) and (d)]. 
            The thick solid lines correspond to the binodals, which delimit the two-phase coexistence regions in the 
            $\phi$-$\chi$ diagrams [(a) and (c)] from below. 
            The dashed lines are the spinodals and the thin horizontal line in panel (a) is the tie-line corresponding
            to the triple point ($\blacktriangle$) found within BCA.
            Representative values for the solubility contrasts per $k_BT$, $(f_+,f_-)=(3,26)$, have been chosen.
            The chemical potential $\mu_Ik_BT$ of the salt corresponds to an ionic strength 
            $\widetilde{I}_c=\widetilde{\rho}_\pm=10\,\m{mM}$ at the critical point with composition 
            $\phi=\phi_c\approx\frac{1}{2}$.
            The weak influence of the salt on the phase diagram within LDA leads to curves in panels (c) and (d) which 
            are, on the present scale, almost (but not quite) symmetric with respect to $\phi=\frac{1}{2}$ and 
            $\mu_\phi=0$, respectively.
            Whereas the LDA [(c) and (d)], in agreement with the experimental evidence, exhibits a single critical point 
            ($\bullet$, $\phi_{c,1}\approx\frac{1}{2},\chi_{c,1}\approx2$), which slightly shifts upon changing the ionic
            strength (see Fig.~\ref{fig:3}), the standard BCA [(a) and (b)], in contrast to the available experimental observations, 
            leads to a second critical point ($\bullet$, $\phi_{c,2}\approx0.1,\chi_{c,2}\approx2.1$) as well as to a 
            triple point ($\blacktriangle$).}
   \label{fig:1}
\end{figure*}

Note that by using the square-gradient form of Eq.~\Eq{df} we implicitly assume that the interactions are short-ranged 
\cite{Evans1979}, i.e., van der Waals forces are not taken into account.
Moreover, layering due to packing effects close to walls is also not accounted for by square-gradient theories.
Nonetheless such a description provides reliable results at mesoscopic scales \cite{Evans1979}. 
Finally, the ionic strength is assumed to be sufficiently low so that one can neglect short-ranged ion-ion 
interactions.
Therefore the ions interact with each other only via the electrostatic field. 
Accordingly, the expression for $\omega^{(\pm)}_\m{ion}$ does not contain additional Flory-Huggins parameters and there
are no square-gradient terms for $\rho_\pm$.
However, these features of the simple functional in Eq.~\Eq{df} are not expected to lessen the main conclusions of our
study, which is devoted to investigate the \emph{kind of influences} of ions on solvent properties, rather than to 
construct models with quantitative predictive power.


\subsection{\label{Sec:Ionsolvent}Ion-solvent interaction}

In Eq.~\Eq{df} the ion-solvent interaction is described, within a local density approximation (LDA), by a solvent-induced 
ion potential, $V_\pm(\phi)k_BT$.
The bilinear coupling approximation (BCA) used in previous investigations (see, e.g., Refs.~\cite{Onuki2004,Onuki2006,%
Okamoto2010,Okamoto2011,Nabutovskii1980,Nabutovskii1985,Nabutovskii1986,BenYaakov2009,Samin2011a}) 
corresponds to the choice $V^\m{(BCA)}_\pm(\phi) := f_\pm\phi$, where $f_\pm k_BT=(f_{\pm A}-f_{\pm B})k_BT$ is the 
\emph{difference} between the bulk solvation free energies of a $\pm$-ion in solvents consisting purely of component $A$,
$f_{\pm A}k_BT$, and purely of component $B$, $f_{\pm B}k_BT$.
The solubility contrasts $f_\pm k_BT$ are also known as Gibbs free energies of transfer.
In this context the only relevant parameters are the two differences $f_\pm=f_{\pm A}-f_{\pm B}$ because the other two 
independent quantities $f_{\pm A}+f_{\pm B}$ can be absorbed as shifts in the definition of the chemical potentials 
$\mu_\pm k_BT$ of the ions.
For bulk systems the BCA is identical to the random phase approximation (RPA) \cite{Hansen1986}, which is expected to be
reliable only if the coupling strengths are much smaller than the thermal energy, i.e., $|f_\pm|\ll1$.
However, for electrolyte solutions, this condition is in general not fulfilled.
Instead, the Gibbs free energies of transfer between two liquids are usually of the order of some $10k_BT$ 
\cite{Marcus1983,Inerowicz1994}.

Figures~\ref{fig:1}(a) and \ref{fig:1}(b) display the bulk phase diagram for a constant chemical potential (per $k_BT$) 
$\mu_I:=\mu_++\mu_-$ of added salt obtained within BCA for the representative values $f_+=3, f_-=26$.
This choice is similar to the Gibbs free energies of transfer for potassium chloride ($\m{KCl}$) from water to acetone:
$f_+=2, f_-=23$ \cite{Marcus1983}.
The condition of local charge neutrality $\rho_+=\rho_-=:I$ in the bulk implies that the ionic strength $I$ depends on 
the chemical potentials $\mu_\pm k_BT$ of the ions only via the sum $\mu_++\mu_-=\mu_I$.
For given uniform composition $\phi$ and ionic chemical potential $\mu_I$ the Euler-Lagrange equation of $\Omega$ in 
Eq.~\Eq{df} with respect to uniform ion densities $\rho_\pm = I$ can be used to express the bulk ionic strength as 
\begin{equation}
   I_\m{bulk}(\phi,\mu_I)=\exp\Big(\frac{1}{2}(\mu_I-V_+(\phi)-V_-(\phi))\Big).
   \label{eq:I}
\end{equation}
Within the present model, $I_\m{bulk}$ is \emph{independent} of the Flory-Huggins parameter $\chi$, i.e., 
it depends on the temperature $T$ only via the normalizations of $\mu_I$ and $V_\pm$, which are defined in units of $k_BT$.
In Fig.~\ref{fig:1} the chemical potential $\mu_I$ of the salt is fixed such that the solvent composition 
$\phi=\frac{1}{2}$ leads to an ionic strength $\widetilde{I}=I\widetilde{a}^{-3}=10\,\m{mM}\approx0.006\,\m{nm^{-3}}$, 
where here and in the following we choose the length scale $\widetilde{a}=2\,\hbox{\AA}$.
Due to the absence of gradients, electric fields, and surfaces, the bulk phase diagram is determined by the 
first, third, and fourth term on the right-hand side of Eq.~\Eq{df}.
The theoretically predicted occurrence of two critical points ($\bullet$) as well as of a triple point
($\blacktriangle$) is not supported by experimental evidence, which signals the breakdown of BCA for such large 
parameters $f_\pm$.
Whereas for most systems it is experimentally difficult to preclude the occurrence of such a second critical point or 
triple point, the experimental resolution is yet sufficiently high to exclude these features to occur visibly to the 
extent as predicted by the BCA (Figs.~\ref{fig:1}(a) and (b)).

A more appropriate approximation for the solvent-induced ion potential $V_\pm(\phi)k_BT$, which is derived in 
Appendix~\ref{AppA}, is given by (see also Fig.~\ref{fig:2}(a))
\begin{equation}
   V_\pm(\phi)=-\ln(1-\phi(1-\exp(-f_\pm))).
   \label{eq:V}
\end{equation}
\begin{figure}[!t]
   \includegraphics[width=8cm]{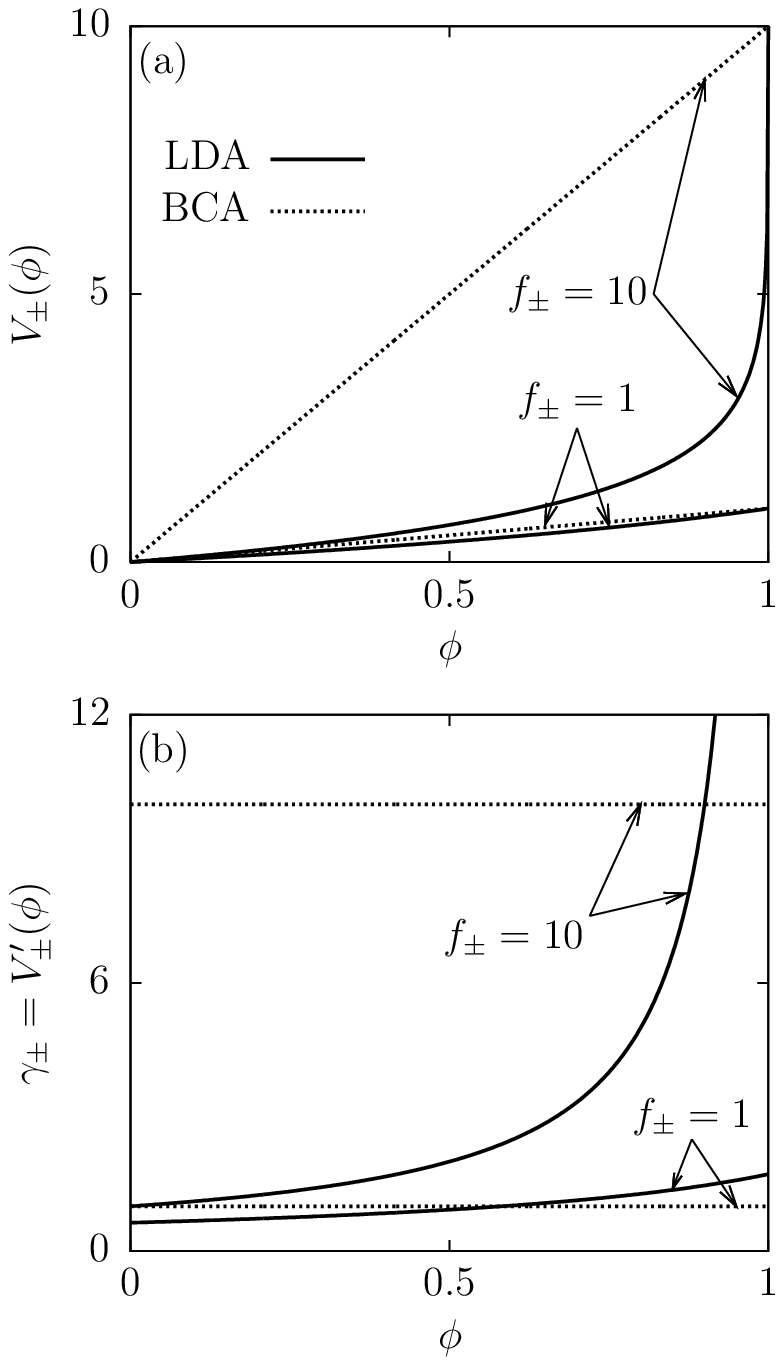}
   \caption{Comparison of the solvent-induced ion potential $V_\pm(\phi)$ [(a)] and its derivative 
            $\gamma_\pm=V'_\pm(\phi)$ [(b)] within LDA and BCA for ion solubility contrast $f_\pm$ (see the main text).
            For small values of $f_\pm$ (see the case $f_\pm=1$) the differences between LDA and BCA are small.
            For large values of $f_\pm$ (see the case $f_\pm=10$) $V_\pm(\phi)$ and $\gamma_\pm=V'_\pm(\phi)$ become
            large at solvent compositions $\phi\approx0.5$ within BCA whereas they remain small within LDA.
            Within LDA $V'_\pm(\phi=0)=1-\exp(-f_\pm)$ and $V'_\pm(\phi=1)=\exp(f_\pm)-1$, while within BCA
            $V'_\pm(\phi)=f_\pm$.}
   \label{fig:2}
\end{figure}
For $|f_\pm| \ll 1$ this expression reduces to the correct asymptotic expression 
$V_\pm(\phi) \simeq V^\m{(BCA)}_\pm(\phi)$.
In the limit $f_\pm\to\infty$, i.e., if ions are insoluble in component $A$, $V_\pm(\phi)\simeq-\ln(1-\phi)$, which
corresponds to the free energy of the ions dissolving entirely in component $B$ only, which has the volume fraction 
$1-\phi$.
Similarly, in the limit $f_\pm\to-\infty$, i.e., if the ions are insoluble in component $B$, 
$V_\pm(\phi)\simeq f_\pm-\ln\phi$, which is the free energy of the ions dissolving entirely in component $A$ only, 
which has the volume fraction $\phi$ and for which the solvation free energy is $f_\pm$.
For the same set of parameters as in Figs.~\ref{fig:1}(a) and (b), Figs.~\ref{fig:1}(c) and (d) display the phase 
diagram within LDA.
In agreement with experimental observations, within LDA only a single critical point ($\bullet$) occurs (see, e.g., the
closed loop-binodals in Ref.~\cite{Sadakane2011} with only one lower critical demixing point in the presence of
an antagonistic salt, i.e., with $f_+$ and $f_-$ having opposite signs).
Hence one can conclude that the standard BCA, i.e., $V^\m{(BCA)}(\phi)=f_\pm\phi$, introduces artifacts for too large  
ion-solvent couplings, $|f_\pm|\gg1$, which are absent within the LDA proposed in Eq.~\Eq{V}.
Note that within LDA the salt has such a weak influence on the phase diagram that the curves in Figs.~\ref{fig:1}(c) 
and (d) are, on that scale, almost (but not quite) symmetric with respect to $\phi=\frac{1}{2}$ and $\mu_\phi=0$, 
respectively, whereas within BCA the influence of salt on the phase diagram is so strong that the curves
in Figs.~\ref{fig:1}(a) and (b) are highly asymmetric.

If $\phi$ deviates slightly from a certain composition $\phi_0\in[0,1]$ one has 
$V_\pm(\phi)\simeq V_\pm(\phi_0)+\gamma_\pm(\phi-\phi_0)$ with the effective coupling strengths
$\dps \gamma_\pm:=V'_\pm(\phi_0)=\frac{1-\exp(-f_\pm)}{1-\phi_0(1-\exp(-f_\pm))}\in
\Big[-\frac{1}{\phi_0},\frac{1}{1-\phi_0}\Big]$ instead of $f_\pm$ as in BCA.
For, e.g., $\phi_0=1/2$ one finds $\gamma_\pm=2\tanh(f_\pm/2)\in[-2,2]$, i.e., the use of BCA, which corresponds
to $\gamma_\pm \approx f_\pm$, is justified only for Gibbs free energies of transfer per $k_BT$, $f_\pm$, not larger
than $2$ (see Fig.~\ref{fig:2}(b)).
However, in previous investigations BCA has been used even for large values of $|f_\pm|$
\cite{Onuki2004,Onuki2006,Okamoto2010,Okamoto2011,BenYaakov2009,Samin2011a}.


\section{\label{Sec:Bulk}Bulk systems}

Bulk properties such as the phase diagram and partial structure factors are determined routinely in order to characterize
experimentally the behavior of fluids.
The available experimental data offer the possibility to assess the quality of the proposed LDA (see the previous 
Sec.~\ref{Sec:Model}) with respect to predictions of bulk properties and to compare it with the frequently used BCA.

\subsection{\label{Sec:Critical}Critical point}

\begin{figure}[!t]
   \includegraphics[width=8cm]{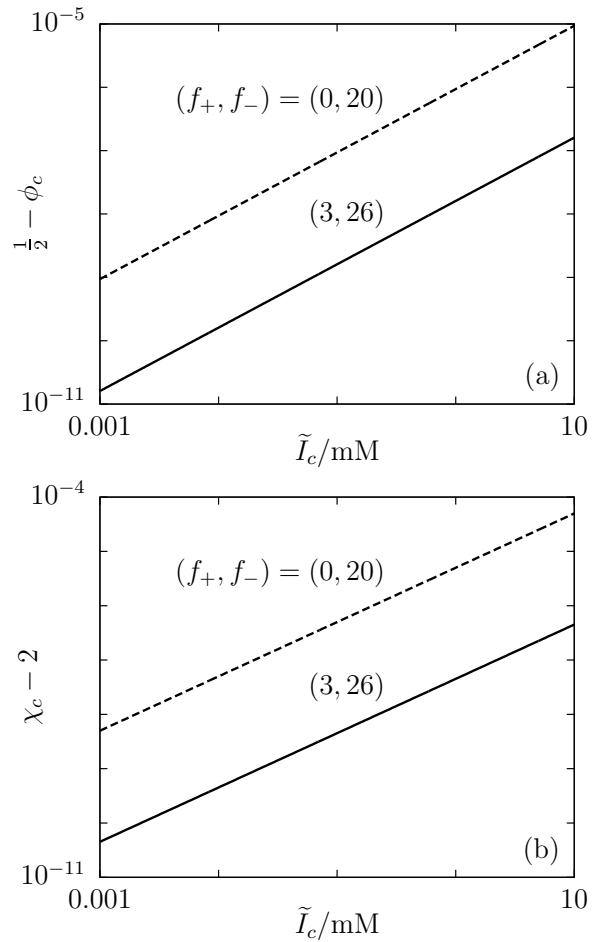}
   \caption{Variation of the critical volume fraction $\phi_c$ [(a)] and the critical Flory-Huggins parameter
            $\chi_c$ [(b)] as function of the ionic strength $\widetilde{I}_c=I_c\widetilde{a}^{-3}$ at the critical 
            point for two representative sets of solubility contrasts: $(f_+,f_-)=(3,26)$ and $(0,20)$.
            These results show that both $\frac{1}{2}-\phi_c$ and $\chi_c-2$ depend linearly on $I_c$ and that 
            there are no quantitatively significant shifts of the critical point upon varying
            the ionic strength within experimentally reasonable ranges.
            On this scale, the phase diagrams for $(f_+,f_-)=(3,26)$ (see Figs.~\ref{fig:1}(c) and (d)) and for 
            $(f_+,f_-)=(0,20)$ are almost indistinguishable.
            Note that $\phi_c(\widetilde{I}_c\to0)=\frac{1}{2}$ and $\chi_c(\widetilde{I}_c\to0)=2$.}
   \label{fig:3}
\end{figure}

There is experimental evidence \cite{Seah1993} that in the phase diagram of a binary liquid mixture the critical point
shifts upon adding salt.
The direction as well as the magnitude of the shift depend on the materials properties of the binary liquid 
mixture and of the ions.
Due to the relations $\rho_A=\phi,\rho_B=1-\phi,\rho_+=\rho_-=I$ the bulk system, which comprises four particle species,
is de facto a binary mixture, characterized by $\mu_\phi$, $\mu_I=\mu_++\mu_-$, and $\chi$ (i.e., $T$).
Hence in this three-dimensional space of thermodynamic variables there is a sheet of first-order demixing phase 
transitions $(\mu^\m{(demix)}_\phi(\mu_I,\chi),\mu_I,\chi)$ bounded by a line of critical points 
$(\mu^\m{(crit)}_\phi(\mu_I), \mu_I, \chi^\m{(crit)}(\mu_I))$ which translates into 
$(\phi_c,I_c,\chi_c)\equiv(\phi^\m{(crit)}(\mu_I),I^\m{(crit)}(\mu_I),\chi^\m{(crit)}(\mu_I))$.
For a given chemical potential $\mu_Ik_BT$ the critical point $(\phi_c,I_c,\chi_c)$ is determined as the minimum of 
the Flory-Huggins parameter $\chi_s(\phi,I_\m{bulk}(\phi,\mu_I))$ at the spinodal as a function of $\phi$ for 
constant $\mu_I$.
The spinodal is defined by the set of points $(\phi,I,\chi_s(\phi,I))$ in the bulk phase diagram for which points 
$(\phi,I,\chi)$ with $\chi>\chi_s(\phi,I)$ exhibit no longer at least a local minimum of the density functional 
Eq.~\Eq{df} (see the dashed lines in Figs.~\ref{fig:1}(a) and (c)).
Accordingly, at the spinodal the Hessian matrix of the bulk grand potential density $\Omega(\phi,I)/V$ corresponding
to Eq.~\Eq{df} has a zero eigenvalue.
This condition leads to 
\begin{eqnarray}
   &&
   \chi_s(\phi,I) = \frac{1}{2}\bigg(\frac{1}{\phi}+\frac{1}{1-\phi} + \\
   \label{eq:chis}
   &&
   \phantom{MM}
   \left(V_+''(\phi)+V_-''(\phi) - \frac{1}{2}(V_+'(\phi)+V_-'(\phi))^2\right)I\bigg). \nonumber
\end{eqnarray}
By inverting the relation $I_c=I^\m{(crit)}(\mu_I)$ one obtains $\mu_I=\mu^\m{(crit)}_I(I_c)$.
Figure~\ref{fig:3} displays the variation of (a) the critical volume fraction 
$\phi_c=\phi^\m{(crit)}(\mu^\m{(crit)}_I(I_c))$ and (b) the critical Flory-Huggins parameter 
$\chi_c=\chi^\m{(crit)}(\mu^\m{(crit)}_I(I_c))$ as functions of the ionic strength 
$\widetilde{I}_c=I_c\widetilde{a}^{-3}$ at the critical point within the present LDA model.
Without added salt ($I=0$) one obtains the critical point $(\phi_c,\chi_c)=(\frac{1}{2},2)$ of the pure solvent. 
For the given choice of the parameters $(f_+,f_-)$ and for small $I_c$ the critical composition $\phi_c$ decreases and 
the critical Flory-Huggins parameter $\chi_c$ increases linearly upon increasing the ionic strength $I_c$.
For small $I_c$ the asymptotically linear dependence of the critical point $(\phi_c,\chi_c)$ on the ionic 
strength $I_c$ is in agreement with experimental evidence \cite{Seah1993,Eckfeldt1943}.
However, the magnitudes of these shifts are tiny, even for large differences in the solubility contrasts, e.g.,
$(f_+,f_-)=(0,20)$, to the effect that the bulk phase diagrams are almost indistinguishable within experimentally 
relevant ranges of ionic strengths $\widetilde{I}_c\lesssim10\,\m{mM}$.
Although for the parameters used in Fig.~\ref{fig:3} the shifts of the critical point $(\phi_c,\chi_c)$ as predicted 
within the BCA are up to three orders of magnitude larger than within the LDA, the effect is still small.
Within the range of ionic strengths considered in Fig.~\ref{fig:3} the experimentally observed critical point shifts 
are also small \cite{Seah1993}. 
However, significant shifts of the critical temperature $T_c$ have been detected for large ionic strengths 
$\widetilde{I}_c\gg100\,\m{mM}$ \cite{Seah1993}.
But for such large ionic strengths the model in Eq.~\Eq{df} is not expected to be applicable, because it neglects 
short-ranged ion-ion interactions.


\subsection{\label{Sec:Correlations}Correlation functions} 

The bulk structure of fluids, which is experimentally accessible by X-ray and neutron scattering, provides 
information complementary to those which follow from the bulk phase behavior.
Hence it provides additional opportunities to assess the quality of the LDA.
Here we consider a spatially uniform equilibrium state $(\phi,I,\chi)$ in the one-phase region of the phase diagram (see 
Fig.~\ref{fig:1}(c)), which minimizes the density functional $\Omega$ in Eq.~\Eq{df} in the absence of surfaces, i.e.,
without the last term therein.
The corresponding two-point correlation functions $G_{ij}(r)=\rho_i\rho_jh_{ij}(r)=\rho_i\rho_j(g_{ij}(r)-1), 
i,j\in\{\phi,+,-\},\rho_\phi:=\phi,\rho_\pm=I$, are obtained from $G_{ij}(r)=G_{ij}(\vec{r},\vec{0}), r=|\vec{r}|,$ with 
the inverse $\dps G^{-1}_{ij}(\vec{r},\vec{r'})=\frac{\delta^2\Omega}{\delta\rho_i(\vec{r})\delta\rho_j(\vec{r'})}$, 
where $\dps\sum_j\Int{\mathcal{V}}{3}{r'}G^{-1}_{ij}(\vec{r},\vec{r'})G_{jk}(\vec{r'},\vec{r''}) = 
\delta_{ik}\delta(\vec{r}-\vec{r''})$.
The three-dimensional Fourier transforms 
$\dps\widehat{G}_{ij}(k):=\frac{4\pi}{k}\int\limits_0^\infty\!\d r\;rG_{ij}(r)\sin(kr)$ with dimensionless $k$, 
which are proportional to the partial structure factors \cite{Hansen1986}, are given by
\begin{widetext}
\begin{eqnarray}
   \widehat{G}_{\phi\phi}(k) & = & 
   \frac{1}{L(k)}\left(k^2+\kappa^2\right), 
   \nonumber\\
   \widehat{G}_{\phi\pm}(k)  & = &  
   -\frac{I}{L(k)}\left(V_\pm'(\phi)k^2+\frac{\kappa^2}{2}(V_+'(\phi)+V_-'(\phi))\right), 
   \nonumber\\
   \widehat{G}_{\pm\pm}(k) & = & 
   \frac{I}{L(k)}\left(\left(\frac{1}{\phi}+\frac{1}{1-\phi}-2\chi+I(V_+''(\phi)+V_-''(\phi))+
   \frac{\chi}{3}k^2\right)\left(k^2+\frac{\kappa^2}{2}\right) - V_\mp'(\phi)^2Ik^2\right),
   \nonumber\\
   \widehat{G}_{\pm\mp}(k) & = & 
   \frac{I}{L(k)}\left(\left(\frac{1}{\phi}+\frac{1}{1-\phi}-2\chi+I(V_+''(\phi)+V_-''(\phi))+
   \frac{\chi}{3}k^2\right)\frac{\kappa^2}{2} + V_+'(\phi)V_-'(\phi)^2Ik^2\right)
   \label{eq:G}
\end{eqnarray}
\end{widetext}
with 
\begin{equation}
   \kappa^2:=\frac{8\pi\ell_BI}{\epsilon(\phi)}
   \label{eq:kappa2}
\end{equation}
as the square of the inverse Debye length and the denominator (see Eq.~\Eq{chis})
\begin{eqnarray}
   L(k) 
   & := & 
   (k^2+\kappa^2)\left(\frac{\chi}{3}k^2 + 2(\chi_s(\phi;I)-\chi)\right) 
   \nonumber\\
   & &
   -\frac{I}{2}(V_+'(\phi)-V_-'(\phi))^2k^2.
   \label{eq:denom}
\end{eqnarray}
Note that $V'_\pm(\phi)=0$ leads to $\widehat{G}_{\phi\pm}(k)=0$, i.e., as expected, the fluctuations of the solvent 
composition and of the ion densities are uncorrelated in the absence of ion-solvent interactions.

Due to the constraint $\rho_A+\rho_B=1$, the correlation functions $\widehat{G}_{AA}(k)$, $\widehat{G}_{AB}(k)$, 
and $\widehat{G}_{BB}(k)$ of the number density fluctuations of the $A$ and $B$ particles are related to the correlation 
function $\widehat{G}_{\phi\phi}(k)$ by 
$\widehat{G}_{AA}(k) = -\widehat{G}_{AB}(k) = \widehat{G}_{BB}(k) = \widehat{G}_{\phi\phi}(k)$.
In the following we refer to $\widehat{G}_{\phi\phi}(k)$ as the solvent structure factor.
It can be written in the form
\begin{equation}
   \widehat{G}_{\phi\phi}(k) = 
   \frac{\dps \widehat{G}_{\phi\phi}(0)}{\dps 1 + (\lambda k)^2\left(1 - \frac{g^2}{1 + (k/\kappa)^2}\right)}
   \label{eq:gphiphi}
\end{equation}
with 
\begin{equation}
   \lambda:=\sqrt{\frac{\chi}{6(\chi_s(\phi,I)-\chi)}}
   \label{eq:lambda}
\end{equation}
and 
\begin{equation}
   g^2:=\frac{3(\Delta\gamma)^2\epsilon(\phi)}{16\pi\ell_B\chi},
   \label{eq:g2}
\end{equation}
where $\Delta\gamma:=\gamma_+-\gamma_-=V_+'(\phi)-V_-'(\phi)$.
The isothermal compressibility, which is proportional to $\widehat{G}_{\phi\phi}(0)=(2(\chi_s(\phi,I)-\chi))^{-1}$,
diverges $\propto|\chi-\chi_c|^{-\gamma}$ upon approaching the critical point $(\phi_c,I_c,\chi_c)$.
As expected within the present mean-field theory, one finds the classical critical exponent $\gamma=1$ instead of 
$\gamma\approx1.24$ for the Ising universality class \cite{Pelissetto2002}.
For a state point $(\phi,I,\chi)$ in the bulk phase diagram (see Figs.~\ref{fig:1}(a) and (c)) the length $\lambda$ 
is an (inverse) measure of the deviation of $\chi$ from its value $\chi_s(\phi,I)$ at the spinodal.
Equation~\Eq{gphiphi} has already been derived in Ref.~\cite{Onuki2004} within BCA, which corresponds to the linear 
approximation $\Delta\gamma\approx f_+-f_-$.
For  $|g|\leq 1$ in Eq.~\Eq{gphiphi} the solvent structure factor $\widehat{G}_{\phi\phi}(k)$ is a monotonically 
decreasing function of the wave number $k$, whereas for $|g| > 1$ at $k_\m{max} = \kappa\sqrt{|g|-1}$ a maximum 
$\dps \widehat{G}_{\phi\phi}(k_\m{max}) = \frac{\widehat{G}_{\phi\phi}(0)}{1 - (\kappa\lambda)^2(|g|-1)^2}$ occurs.
Hence, if $|g|>1$, $\widehat{G}_{\phi\phi}(k_\m{max})$ diverges as function of $\chi$ at $\lambda=\lambda_\m{unstable}=(\kappa(|g|-1))^{-1}$,
i.e., the spatially uniform bulk state becomes unstable upon approaching the critical point.
Note that in the limits $|g|\to0$ (no ion-solvent coupling) or $\kappa\to0$ (no salt) Eq.~\Eq{gphiphi} leads to the
Ornstein-Zernike-like solvent structure factor $\widehat{G}_{\phi\phi}(k)=\widehat{G}_{\phi\phi}(0)/(1+(\lambda k)^2)$.
In this case $\lambda$ can be identified with the bulk correlation length.

Experimental reports of uniform bulk states close to the critical point of water+2,6-dimethylpyridine
mixtures with $\m{KBr}$, $\m{KCl}$, and $\m{Mg(NO_3)_2}$ (see Ref.~\cite{Nellen2011}) as well as  
distributions of neutron scattering intensities of water+3-methylpyridine with $\m{LiCl}$, 
$\m{NaCl}$, $\m{KCl}$, $\m{NaBr}$, and $\m{MgSO_4}$, which vary monotonically as function of $k$ (see 
Refs.~\cite{Sadakane2006,Sadakane2007a}), indicate that in these systems one has $|g|<1$.
Within the present LDA this latter relation is expected to be fulfilled: Close to the critical point 
$(\phi_c,\chi_c)\approx(\frac{1}{2},2)$ (see Fig.~\ref{fig:3}) of, e.g., the widely studied binary 
liquid mixture of 3-methylpyridine (component $A$, $\epsilon_A=10$) and water (component $B$, $\epsilon_B=80$) with a
lower critical demixing point at $T_c\approx316\,\m{K}$, i.e., $\ell_B\widetilde{a}\approx529\,\hbox{\AA}$, one 
obtains $|g|<0.3$ independent of the type of salt, because $|\Delta\gamma|\lesssim4$ (see the last paragraph in 
Subsec.~\ref{Sec:Ionsolvent}).
However, within BCA, i.e., for $\Delta\gamma\approx f_+-f_-$ with typically $|f_+-f_-|\gg1$ 
\cite{Marcus1983,Inerowicz1994}, one has to expect $|g|\gg1$, which, according to the above reasoning, is in sharp 
contrast to the available experimental results.
We mention that experimental reports \cite{Sadakane2007b,Sadakane2009,Sadakane2011} of ``periodic structures'' in heavy 
water+3-methylpyridine mixtures with sodium tetraphenylborate ($\m{NaBPh_4}$) cannot, however, be expected
to find a consistent interpretation in terms of a local ion solvation model as given in Eq.~\Eq{df}, neither within BCA
nor within the present LDA, because the anions ($\m{[BPh_4]^-}$) are much larger than the solvent particles, such that
in these systems the ion size is expected to be relevant.

The charge-charge structure factor $\dps S_{ZZ}(k)=(\widehat{G}_{\pm\pm}(k)-\widehat{G}_{\pm\mp}(k))/I$ \cite{Hansen1986},
which measures correlations of fluctuations $Z$ of the local charge density around $\rho_+-\rho_-=0$, is obtained 
by inserting the expressions for $\widehat{G}_{\pm\pm}(k)$ and $\widehat{G}_{\pm\mp}(k)$ from Eq.~\Eq{G}:
\begin{equation}
   S_{ZZ}(k) = k^2\frac{\dps\frac{\chi}{3}k^2 + 2(\chi_s(\phi,I)-\chi)}{L(k)}.
   \label{SQQ}
\end{equation}
The asymptotic behavior $S_{ZZ}(k\to0)\simeq (k/\kappa)^2$ is the signature for perfect screening \cite{Hansen1986}.
Further, the case $\Delta\gamma=0$ corresponds to the Debye-H\"{u}ckel limit $S_{ZZ}(k) = k^2/(k^2+\kappa^2)$.

\begin{figure}[!t]
   \includegraphics[width=8cm]{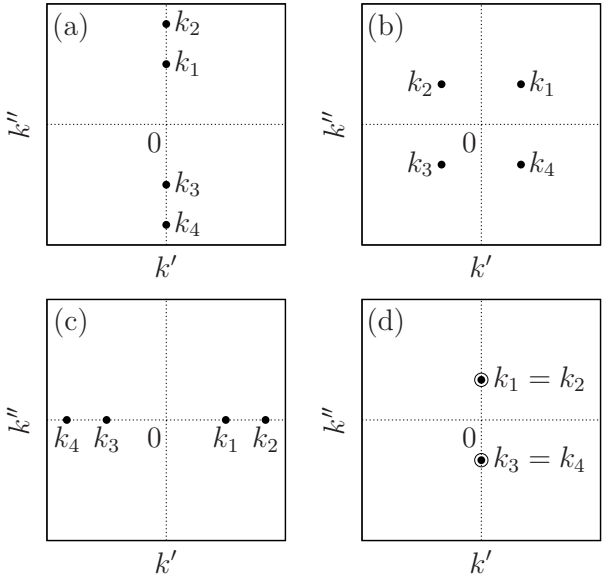}
   \caption{Poles $k_1,\dots,k_4$ of the Fourier transform $\widehat{G}_{ij}(k)$ of the two-point correlation functions 
            $G_{ij}(r)$ in the complex plane $k=k'+ik''\in\mathbb{C}$, which correspond to the roots of the denominator
            $L(k)$ (see Eq.~\Eq{denom}).
            According to the analytic structure of $L(k)$ (see the main text) only the three distinct situations shown in 
            panels (a)--(c) can occur.
            Purely imaginary poles [(a)] correspond to a monotonic decay of $G_{ij}(r\to\infty)$ whereas a pole structure
            as in panel (b) ($|k_\nu|$ all equal) corresponds to an oscillatory decay of $G_{ij}(r\to\infty)$.
            Purely real poles [(c)] indicate an unstable bulk state, which does not occur in the one-phase region of the
            phase diagram in Fig.~\ref{fig:1}(c).
            The merging of two poles on the imaginary axis [(d)] corresponds to a Kirkwood crossover point.}
   \label{fig:4}
\end{figure}

The asymptotic behavior of the correlation function 
$\dps G_{ij}(r)=\frac{1}{2\pi^2r}\int\limits_0^\infty\!\d k\;k\widehat{G}_{ij}(k)\sin(kr)$ can be inferred from a pole
analysis of $\widehat{G}_{ij}(k)$, which amounts to determine the roots of the denominator $L(k)$ defined in 
Eq.~\Eq{denom} \cite{Evans1993,Evans1994}.
Since $L(k)$ is a polynomial in $k$ of degree four it has four and only four complex roots 
$k_\nu=k'_\nu+ik''_\nu,k'_\nu=\Re(k_\nu),k''_\nu=\Im(k_\nu),\nu\in\{1,\dots,4\}$.
Due to the actual structure of $L(k)$ there are constraints on the locations of the four roots $k_\nu$ in the complex
plane.
If $L(k=k_\nu)$ vanishes this holds also for $k=k_\nu^*$, because $L(k)$ has real coefficients.
Moreover, if $L(k=k_\nu)$ vanishes this also holds for $k=-k_\nu$, because $L(k)$ is a polynomial in $k^2$.
Accordingly this is also true for $k=-k_\nu^*$. 
This implies the root structure shown in Fig.~\ref{fig:4}.
Three distinct situations can occur. 
For purely imaginary roots given by $\{k_1=ik_1'',k_2=ik_2'',k_3=-k_1,k_4=-k_2\}$ with $0<k_1''<k_2''$ (see 
Fig.~\ref{fig:4}(a)) the asymptotic decay of the two-point correlation functions $G_{ij}(r\to\infty)$ is monotonic 
$\propto\exp(-k_1''r)/r$.
For complex roots $\{k_1=k_1'+ik_1'',k_2=-k_1^*,k_3=-k_1,k_4=k_1^*\}$ with $k_1',k_1''>0$ (see Fig.~\ref{fig:4}(b)) 
the two-point correlation functions $G_{ij}(r)$ vary asymptotically 
$\propto\sin(k_1'r+\m{const})\exp(-k_1''r)/r$ giving rise to a damped oscillatory decay.
Finally, purely real roots $\{k_1=k_1',k_2=k_2',k_3=-k_1,k_4=-k_2\}$ with $0<k_1'<k_2'$ (see Fig.~\ref{fig:4}(c)) 
indicate an unstable bulk state, i.e., the corresponding point in the phase diagram is located in between the spinodals.
The exponential decay of the two-point correlation functions (whether monotonically or oscillatory) is 
consistent with the short range of the interactions implied by taking a gradient expansion in Eq.~\Eq{df}.

\begin{figure*}[!t]
   \includegraphics[width=16cm]{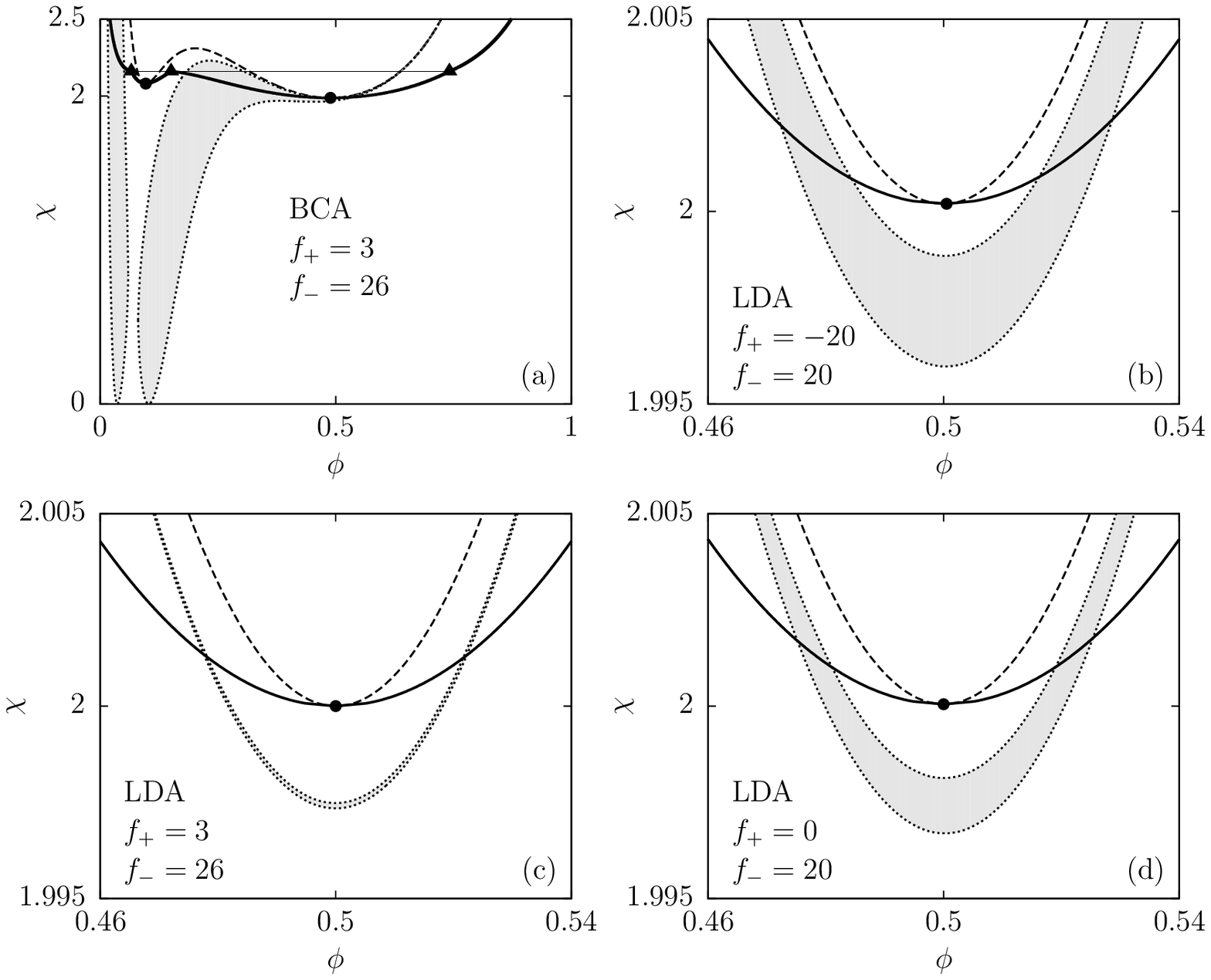}
   \caption{Phase diagrams as in Fig.~\ref{fig:1} with Kirkwood crossover lines (dotted lines) within the bilinear
            coupling approximation (BCA) [(a)] and the local density approximation (LDA) [(b)--(d)].
            The parameters correspond to the binary liquid mixture water+3-methylpyridine ($\epsilon_A=10$, 
            $\epsilon_B=80$, $\ell_B\widetilde{a}=529\,\hbox{\AA}$); for simplicity the temperature dependence of the 
            Bjerrum length $\ell_B$ is ignored. 
            The chemical potential $\mu_Ik_BT$ of the salt is fixed such that at the (slightly shifted) critical point
            with composition $\phi=\phi_c\approx\frac{1}{2}$ there is an ionic strength $\widetilde{I}_c=10\,\m{mM}$.
            Outside the grey regions bounded by the dotted lines the two-point correlation functions exhibit 
            asymptotically a monotonic decay, whereas inside these regions damped oscillatory decays occur.
            Within BCA a large portion of the phase diagram corresponds to oscillatory decay, whereas within LDA this
            occurs only in a narrow band within which the value of the bulk correlation length is close to that of 
            the Debye screening length.
            Note the differences in scales for the axes in (a) and in (b)--(d).}
   \label{fig:5}
\end{figure*}

Thermodynamic states in the bulk phase diagram with monotonically decaying $G_{ij}(r\to\infty)$ are separated from 
states with damped oscillatory decay of $G_{ij}(r\to\infty)$ by so-called Kirkwood crossover lines \cite{Carvalho1994}.
Crossing these lines is associated with the merging of two purely imaginary poles (see Figs.~\ref{fig:4}(a) and (d))
of $\widehat{G}_{ij}(k)$ in the upper (and similarly in the lower) half of the complex plane and with a subsequent 
emergence of a pair of two poles (see Fig.~\ref{fig:4}(b)) with equal imaginary parts and with real parts of equal 
absolute value but of opposite sign \cite{Kirkwood1936}.
In the phase diagrams of Fig.~\ref{fig:5} the Kirkwood crossover lines are denoted by dotted lines and damped oscillatory
decay of $G_{ij}(r\to\infty)$ occurs at state points in the grey area enclosed by the Kirkwood crossover lines.
The parameters correspond to the aforementioned binary liquid mixture water+3-methylpyridine ($\epsilon_A=10$, 
$\epsilon_B=80$, $\ell_B\widetilde{a}=529\,\hbox{\AA}$); for simplicity the temperature dependence of the Bjerrum 
length $\ell_B$ is ignored. 
The chemical potential $\mu_I$ of the salt is fixed such that there is an ionic strength $\widetilde{I}_c=10\,\m{mM}$ at
the (shifted) critical point.
Figures~\ref{fig:5}(a) and (c) correspond to the parameters $(f_+,f_-)=(3,26)$ used in Figs.~\ref{fig:1}(a) (BCA) and (c)
(LDA), respectively.
Figure~\ref{fig:5}(b) refers to the case of a strongly antagonistic salt, $(f_+,f_-)=(-20,20)$, whereas Fig.~\ref{fig:5}(d)
relates to the intermediate case $(f_+,f_-)=(0,20)$.
Within BCA (see Fig.~\ref{fig:5}(a)), the damped oscillatory decay of $G_{ij}(r\to\infty)$ prevails in a large portion of
the phase diagram, and wave lengths of the oscillations as small as the particle size $\widetilde{a}$ can occur at 
state points in the center of the grey area.
However, within the present LDA, damped oscillatory decay of $G_{ij}(r\to\infty)$ is found only in a narrow range of
$\mathcal{O}(\Delta\gamma I)$ for values of $\chi$ around $\dps\chi=\frac{\chi_s(\phi,I)}{1+\kappa^2/6}+
\mathcal{O}((\Delta\gamma)^2I)$, which extends into the one-phase region only in the vicinity of the critical point
(see Figs.~\ref{fig:5}(b)--(d)).

For small wave numbers $k$ the structure factor 
$\dps\widehat{G}_{\phi\phi}(k\ll\kappa)\simeq\frac{\widehat{G}_{\phi\phi}(0)}{1+(\lambda k)^2(1-g^2)}$ (see 
Eq.~\Eq{gphiphi}) takes the Ornstein-Zernike form $\dps\frac{\widehat{G}_{\phi\phi}(0)}{1+(\xi^\m{(OZ)} k)^2}$ with the 
length 
\begin{equation}
   \xi^\m{(OZ)}=\lambda\sqrt{1-g^2} = \sqrt{\frac{\chi(1-g^2)}{6(\chi_s(\phi,I)-\chi)}}
   \label{eq:xiOZ}
\end{equation}
which we shall refer to as the Ornstein-Zernike length.
This length $\xi^\m{(OZ)}$ is determined routinely in scattering experiments by fitting an
Ornstein-Zernike expression to scattered intensities at small momentum transfer \cite{Nellen2011}.
For water+2,6-dimethylpyridine mixtures with $\m{KBr}$, $\m{KCl}$, and $\m{Mg(NO_3)_2}$ (see Ref.~\cite{Nellen2011}) 
it has been found experimentally that the amplitude $\xi^\m{(OZ)}_0$ of $\xi^\m{(OZ)}=\xi^\m{(OZ)}_0|(T-T_c)/T_c|^{-\nu}$
is to a large extent independent of the considered type of salt and ionic strength.
Due to $\xi^\m{(OZ)}_0\propto\sqrt{1-g^2}$ this observation indicates that one has $g^2\ll1$, which, according to the 
arguments given above, is expected within LDA but is not compatible with predictions following from BCA.

The poles $\{k_1,\dots,k_4\}$ of the solvent structure factor $\widehat{G}_{\phi\phi}(k)$ can be expressed in 
terms of the Ornstein-Zernike length $\xi^\m{(OZ)}$ and the inverse Debye length $\kappa$.
For a monotonic decay of $G_{\phi\phi}(r\to\infty)$ one has purely imaginary poles $k_\nu=ik''_\nu$ with
\begin{equation} 
   k_1''= -k_3'' = \sqrt{2(u-\sqrt{u^2-v^2})}
   \label{eq:k1__}
\end{equation}
and
\begin{equation}
   k_2''= -k_4'' = \sqrt{2(u+\sqrt{u^2-v^2})},
   \label{eq:k2__} 
\end{equation}
whereas a damped oscillatory decay of $G_{\phi\phi}(r\to\infty)$ is characterized by the poles at 
$\{k_1=k_1'+ik_1'',k_2=-k_1^*,k_3=-k_1,k_4=k_1^*\}$ with
\begin{equation} 
   k_1'=\sqrt{v-u}, \quad k_1''=\sqrt{v+u},
   \label{eq:k1_}
\end{equation}
where 
\begin{equation}
   u := \frac{\kappa^2(1-g^2)}{4}\Big(1+\frac{1}{(\kappa\xi^\m{(OZ)})^2}\Big)
   \label{eq:defu}
\end{equation}
and
\begin{equation}
   v := \frac{\kappa\sqrt{1-g^2}}{2\xi^\m{(OZ)}}.
   \label{eq:defv}
\end{equation}
Close to the critical point (i.e., for $\xi^\m{(OZ)}\to\infty$) one finds a monotonic decay of $G_{\phi\phi}(r\to\infty)$
with the decay length $1/k_1''\simeq\xi^\m{(OZ)}\propto|\chi-\chi_c|^{-\nu}$ (see Eq.~\Eq{k1__}) with the mean-field 
critical exponent $\nu=\frac{1}{2}$ instead of $\nu\approx0.63$ for the Ising universality class \cite{Pelissetto2002}.
Therefore the electrostatic interactions do not affect the universal critical exponent, but they can influence the 
non-universal critical amplitude $\xi^\m{(OZ)}_0$.

\begin{figure}[!t]
   \includegraphics[width=8cm]{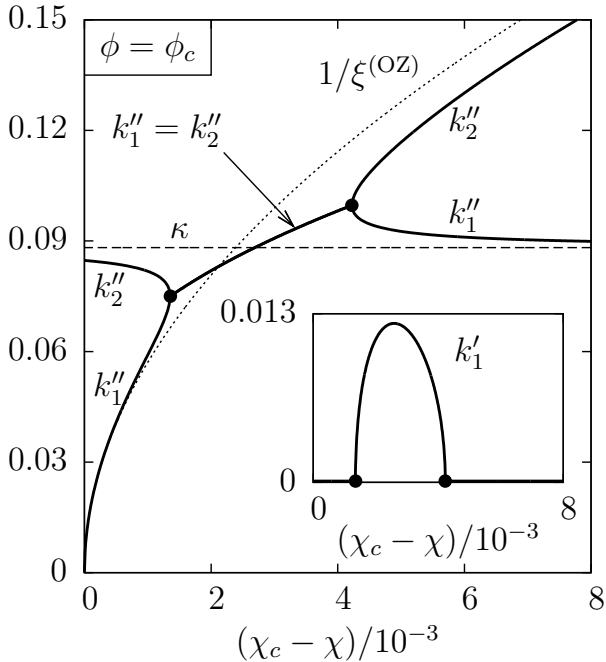}
   \caption{Within LDA real and imaginary parts of the poles $k_\nu=k'_\nu+ik''_\nu,k'_\nu=\Re(k_\nu),
            k''_\nu=\Im(k_\nu),\nu\in\{1,\dots,4\},$ of the Fourier transform $\widehat{G}_{ij}(k)$ of 
            the two-point correlation functions $G_{ij}(r)$ as functions of the deviation $\chi_c-\chi$ from the critical
            point at the critical composition $\phi=\phi_c\approx\frac{1}{2}$ for the parameters corresponding to
            Fig.~\ref{fig:5}(b).
            The four poles $k_1,\dots,k_4$ can be expressed in terms of $(k'_1,k''_1,k''_2)$ (see 
            Eqs.~\Eq{k1__}--\Eq{defv} and Fig.~\ref{fig:4}).
            If $G_{ij}(r\to\infty)$ decays monotonically, the poles of $\widehat{G}_{ij}(k)$ are purely imaginary 
            ($k'_1=0$), giving rise to two branches $k''_1$ and $k''_2$ of positive imaginary parts (see 
            Fig.~\ref{fig:4}(a)).
            If $G_{ij}(r\to\infty)$ decays oscillatorily, there is only one pole of $\widehat{G}_{ij}(k)$ with positive 
            real and imaginary parts ($k''_2=k''_1$) (see Fig.~\ref{fig:4}(b)).
            The merging of the two branches $k''_1$ and $k''_2$ for monotonic asymptotic decay takes place at the 
            Kirkwood crossover points ($\bullet$) (see Fig.~\ref{fig:4}(d)).
            Upon varying $\phi$ these points form the Kirkwood crossover lines (dotted lines in Fig.~\ref{fig:5}).
            For comparison the inverse Debye length $\kappa$ (dashed line) as well as the inverse Ornstein-Zernike
            length $1/\xi^\m{(OZ)}\propto\sqrt{\chi_c-\chi}$ (dotted line, see Eq.~\Eq{xiOZ}) are displayed.
            Within the range of values of $\chi$ leading to an oscillatory decay, depicted by the grey regions in 
            Fig.~\ref{fig:5}, one has $\kappa\approx1/\xi^\m{(OZ)}$.
            Within the range of monotonic decay the decay rate of the leading contribution to $G_{ij}(r\to\infty)$
            is given by $k''_1$ whereas that of the subdominant contribution is $k''_2$.
            For $\chi_c-\chi\leq1.3\times10^{-3}$ the decay rates are $k''_1\approx1/\xi^\m{(OZ)}$ and 
            $k''_2\approx\kappa$, whereas for $\chi_c-\chi\geq4.3\times10^{-3}$ the decay rates are $k''_1\approx\kappa$
            and $k''_2\approx1/\xi^\m{(OZ)}$.} 
   \label{fig:6}
\end{figure}

Figure~\ref{fig:6} displays the real and imaginary parts of the poles $k_\nu$ of $\widehat{G}_{ij}(k)$ in the ranges
$k_\nu'=\Re(k_\nu),k_\nu''=\Im(k_\nu)\geq0$ at the critical composition $\phi=\phi_c\approx\frac{1}{2}$ for the
parameters corresponding to Fig.~\ref{fig:5}(b).
The four poles $\{k_1,\dots,k_4\}$ can be expressed in terms of $(k'_1,k''_1,k''_2)$ (see Eqs.~\Eq{k1__}--\Eq{k1_}).
In Fig.~\ref{fig:6} the two purely imaginary poles $k_1$ and $k_2$ with positive imaginary parts for monotonically 
decaying $G_{ij}(r\to\infty)$ occur as two branches, which merge at the Kirkwood crossover points
($\bullet$).
From Eqs.~\Eq{k1__} and \Eq{k2__} the Kirkwood crossover points ($\bullet$) are characterized by $u=v$ which leads 
to (see Eqs.~\Eq{defu} and \Eq{defv}) $\dps\kappa\xi^\m{(OZ)} = \sqrt{\frac{2}{1\pm g}-1} \approx 1$.
Hence at the Kirkwood crossover points the inverse decay lengths of $G_{ij}(r\to\infty)$ correspond approximately 
to $\kappa$ (dashed line) and $1/\xi^\m{(OZ)}$ (dotted line).
At the critical point ($\chi=\chi_c$, i.e., $k''_1=0$, see Eqs.~\Eq{k1__} and \Eq{defv}), $G_{ij}(r\to\infty)$ decays
as $1/r$ with a subdominant contribution $\propto\exp(-\kappa r\sqrt{1-g^2})/r$ (see Fig.~\ref{fig:6}), i.e., 
as anticipated above, the leading decay at large distances is governed by the vicinity to the critical point, whereas
the ion-solvent coupling manifests itself in the corrections to the leading behavior.
Further away from the critical point the leading contribution decays $\propto\exp(-k_1''r)/r$ with a subdominant contribution 
$\propto\exp(-k_2''r)/r$ (see Fig.~\ref{fig:6}).
The inset of Fig.~\ref{fig:6} displays the absolute value of the real parts $|k'_\nu|=k'_1$ of the poles of 
$\widehat{G}_{ij}(k)$, which is identical to the wave number $k_1'$ (see Eq.~\Eq{k1_}) of the oscillatory part of 
$G_{ij}(r)$ and which is non-zero within the grey region of Fig.~\ref{fig:5}(b).
For the strongly antagonistic salt with $f_+=-20$ and $f_-=20$, in Fig.~\ref{fig:5}(b) the shortest wave length 
of the oscillations is given by $(2\pi/k'_1)_\m{min} \approx 513$ (see the inset in Fig.~\ref{fig:6}).
The corresponding value of $k_1''$ is $\approx0.0873$ so that $(2\pi/k'_1)_\m{min} \approx 45/k''_1$, i.e., 
$G_{ij}(r)\propto\sin(k'_1r+\m{const})\exp(-k''_1r)/r$ decays already within $1/45$ of a period.
In less extreme cases of solubility contrasts $f_\pm$, such as those in Figs.~\ref{fig:5}(c) and (d), the shortest 
wave lengths are even larger.
Therefore, within LDA, the oscillations in $G_{ij}(r)$, if they occur, are not expected to be experimentally detectable.
In contrast, as already mentioned above, within BCA it is possible that the shortest wave lengths of the oscillations 
are of the order of the particle size; such an asymptotic oscillatory decay can be expected to be visible in the pair
distribution function.
However, we are not aware of any experimental reports of Kirkwood crossover lines, which is in line with the results 
obtained within the LDA.


\section{\label{Sec:Interface}Semi-infinite systems bounded by a planar wall}

Interfacial properties such as number density profiles and the excess adsorption are experimentally accessible, albeit
requiring more effort than determining bulk properties.
However, if available, they provide information which is significantly more detailed than the one which can be inferred
from the bulk properties discussed in the previous Sec.~\ref{Sec:Bulk}.
In this respect the simplest and experimentally appealing setting is that of a semi-infinite planar system, on which
we shall focus in the following.
Predictions within the proposed LDA (see Sec.~\ref{Sec:Model}) will be compared with those obtained within the BCA as 
well as with the limited amount of corresponding experimental data which are presently available.
Additional experimental settings are proposed in order to assess the predictive power of the LDA.

\subsection{Profiles} 

According to the fluctuation-dissipation theorem the linear density response of a system to a weak external field
is determined by the two-point correlation functions of the unperturbed system \cite{Hansen1986}.
Therefore, the number density profiles far from a wall exhibit asymptotically the same type of decay towards their 
bulk values, i.e., either monotonically or damped oscillatorily, with the same decay length and periodicity 
as the two-point bulk correlation functions $G_{ij}(r)$.
(Note that this correspondence does no longer hold in the presence of algebraically decaying interaction potentials
\cite{Dietrich1991}, which we do not consider here.)
However, from this argument one cannot draw reliable conclusions concerning their structure close to the wall.
Therefore in this latter range we determine the structure numerically for particular sets of parameters.
Moreover, close to the wall packing effects due to the finite size of the fluid particles lead to layering 
which extends a few particle diameters into the system.
However, this kind of structure is not captured by the present square-gradient model (Eq.~\Eq{df}).

\begin{figure*}[!t]
   \includegraphics[width=16cm]{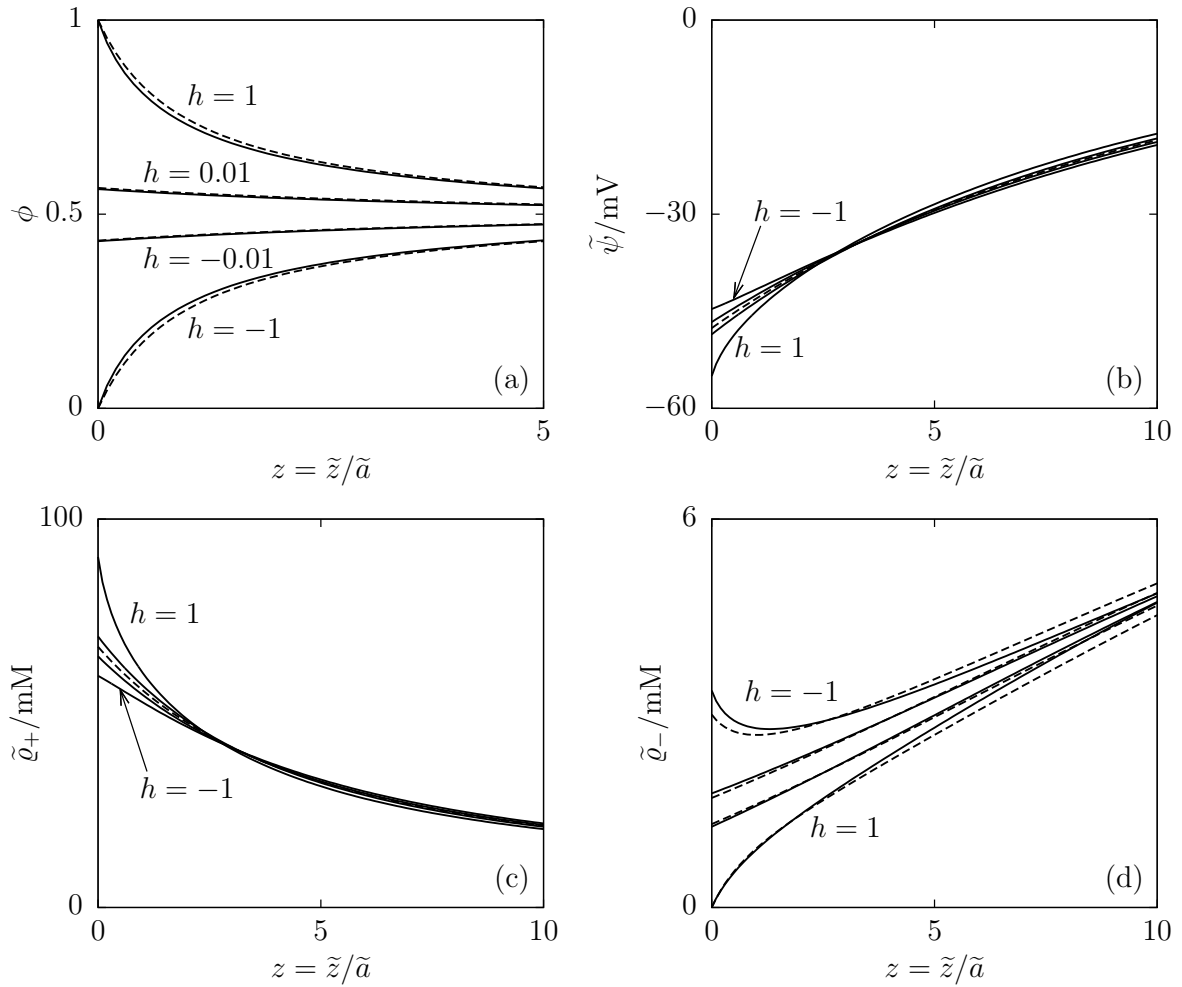}
   \caption{Profiles of the volume fraction $\phi$ of solvent component $A$ [(a)], the electrostatic potential 
            $\widetilde{\psi}$ (with $\widetilde{\psi}(z\to\infty)\to0$) [(b)], the cation number density 
            $\widetilde{\rho}_+$ [(c)], and the anion number density $\widetilde{\rho}_-$ [(d)] in a semi-infinite system 
            bounded by a wall at $z=\widetilde{z}/\widetilde{a}=0$ with surface charge density $\widetilde{\sigma}=
            -1\,\m{\mu C/cm^2}$ and surface field strength $h$.
            These results correspond to Gibbs free energies of transfer $f_+=0,f_-=20$, the bulk volume fraction 
            $\phi_b=0.5$ of solvent component $A$, and the bulk ionic strength $\widetilde{I}=\widetilde{\rho}_
            {\pm b}=10\,\m{mM}$.
            The Flory-Huggins parameter $\chi(T)$ is chosen to correspond to that temperature, for which the bulk 
            correlation length $\xi$ is half of the Debye length $1/\kappa$, which is taken to be temperature 
            independent (see Fig.~\ref{fig:6}).
            For the specified surface fields $h$ the solid lines are the numeric solutions obtained from the density
            functional model in Eq.~\Eq{df} within LDA.
            For reasons of clarity in (b) and (c) the full lines for $h=\pm0.01$ are not designated; they can be 
            nonetheless identified in an obvious way.
            The dashed lines correspond to the approximate profiles $\overline{\phi}(z)$, $\overline{\psi}(z)$, and
            $\overline{\rho}_\pm(z)$ introduced in Eqs.~\Eq{phiapprox}, \Eq{psiapprox}, and \Eq{rhoapprox}, respectively.
            Note that $\overline{\psi}(z)$ and, due to the choice $f_+=0$, $\overline{\rho}_+(z)$ are independent of 
            the magnitude $|h|$; therefore both in (b) and (c) there is only one dashed line.
            For $z>2$ the approximate profiles differ only slightly from the ones obtained by a full numerical 
            minimization.
            Density oscillations close to the wall, which are expected in actual fluids, do not occur, because packing
            effects are not captured by the present square-gradient approach.}             
   \label{fig:7}
\end{figure*}

The solid lines in Fig.~\ref{fig:7} correspond to the composition $\phi(z)$ [(a)], the electrostatic potential
$\widetilde{\psi}(z)=\psi(z)k_BT/e$ with $\epsilon(\phi)\psi'(z)=-4\pi\ell_BD(z)$ and $\psi(z\to\infty)\to0$ [(b)], 
the cation number density $\rho_+(z)$ [(c)], and the anion number density $\rho_-(z)$ [(d)] in a semi-infinite system 
bounded by a wall positioned at $z=0$ with surface charge density $\sigma$ and surface field strength $h$, as obtained
from numerically minimizing the density functional in Eq.~\Eq{df}.
The solvent permittivity is chosen to resemble that of a mixture of 3-methylpyridine (component $A$, $\epsilon_A=10$)
and water (component $B$, $\epsilon_B=80$).
The composition profiles $\phi(z)$ in Fig.~\ref{fig:7}(a) turn out to be monotonic for weak ($h=\pm0.01$) as well as 
for strong surface fields ($h=\pm1$).
Due to the negative surface charge density $\sigma$, the monotonic electrostatic potential profile $\psi(z)$ in
Fig.~\ref{fig:7}(b) is negative with surface potentials $\widetilde{\psi}(0)$ of some tens of $\m{mV}$, which is a 
common order of magnitude \cite{Davis1978}.
Within the range $0\leq z \leq 2$ close to the wall the electrostatic potential $\psi$ becomes less negative upon 
changing the surface field strength from $h=1$ to $h=-1$ due to the increase of the permittivity as a result of the 
increase of the volume fraction $1-\phi$ of component $B$ close to the wall.
Similarly, due to the negative surface charge, close to the wall the number density $\rho_+$ of the cations in 
Fig.~\ref{fig:7}(c) is larger and the number density $\rho_-$ of the anions in Fig.~\ref{fig:7}(d) is smaller than 
in the bulk.
Upon changing the surface field strength from $h=1$ to $h=-1$, close to the wall the number density $\rho_+$ of the 
cations decreases and that of the anions, $\rho_-$, increases.
This feature follows partly from the variation of the electrostatic potential $\psi$.
In addition, for the current choice of parameters the anions dissolve better in component $B$ than in component
$A$ of the solvent ($f_->0$), such that the component $B$ enriched near the surface (see Fig.~\ref{fig:7}(a)) mediates
a certain preference of the anions for the wall.

The dashed lines in Fig.~\ref{fig:7}(a) correspond to the approximate profile
\begin{eqnarray}
   &&
   \overline{\phi}(z) = \phi_b + \frac{C_{GL}}{\sinh((z+z_0)/\xi)}, \nonumber\\
   && 
   C_{GL} := \sign(h)\sqrt{\frac{\chi}{8\xi^2}}
   \label{eq:phiapprox}
\end{eqnarray}
with the extrapolation length $z_0$.
Here and in the following we call $\xi \equiv \xi^\m{(OZ)}$ (see Sec.~\ref{Sec:Correlations}) the bulk correlation length.
As noted in Sec.~\ref{Sec:Correlations}, close to the critical point $1/\xi$ corresponds to the asymptotic decay rate
$k''_1$ of the solvent structure factor $G_{\phi\phi}(r\to\infty)\propto\exp(-r/\xi)/r$. 
The profile $\overline{\phi}(z)$ is the analytic solution of the semi-infinite Ginzburg-Landau 
equation \cite{Lubensky1975} obtained from minimizing $\Omega[\phi,\rho_\pm]$ after expanding Eq.~\Eq{df} up to fourth 
order in $\phi-\phi_b$ and neglecting the ion-solvent coupling, i.e., assuming $f_\pm=0$, which implies $V_\pm(\phi)=0$.
Within Ginzburg-Landau theory the extrapolation length $z_0$ is fixed by the boundary condition 
${\overline{\phi}}\,'(0)=-3h/\chi$.
For $|h|\to\infty$ or $\xi\to\infty$ this leads to $|\overline{\phi}(0)|\to\infty$.
However, within the context of the present study $\overline{\phi}(z)$ is an approximation of the volume fraction of 
solvent component $A$, which is restricted to the interval $[0,1]$.
Hence, we accept the extrapolation length $z_0$ as determined from the boundary condition 
${\overline{\phi}}\,'(0)=-3h/\chi$ only if this leads to $\overline{\phi}(0)\in[0,1]$.
Otherwise the extrapolation length $z_0$ is inferred from the ersatz boundary condition $\overline{\phi}(0)=1$ if 
$h>0$ or $\overline{\phi}(0)=0$ if $h<0$.

The dashed line in Fig.~\ref{fig:7}(b) is the approximate electrostatic potential
\begin{eqnarray}
   &&
   \overline{\psi}(z) = 4\artanh(C_{PB}\exp(-\kappa z)), \nonumber\\
   &&
   C_{PB} := \tanh\Big(\frac{1}{2}\arsinh\Big(\frac{2\pi\ell_B\sigma}{\epsilon(\phi_b)\kappa}\Big)\Big),
   \label{eq:psiapprox}
\end{eqnarray}
which is the analytic solution of the semi-infinite Poisson-Boltzmann equation \cite{Gouy1910} for a uniform 
permittivity $\epsilon(\phi_b)$ and for neglecting the ion-solvent coupling (i.e., for $f_\pm=0$).

Finally, the dashed lines in Figs.~\ref{fig:7}(c) and (d) are the approximate number density profiles of the $\pm$-ions,
\begin{equation}
   \overline{\rho}_\pm(z) = I_b\exp(-(\pm\overline{\psi}(z) + V_\pm(\overline{\phi}(z))) + V_\pm(\phi_b)),
   \label{eq:rhoapprox}
\end{equation}
which correspond to the Boltzmann distributions of non-interacting particles in the external fields due to  the 
approximate electrostatic potential $\overline{\psi}(z)$ and the approximate composition $\overline{\phi}(z)$; 
$I_b=\rho_\pm(z\to\infty)$.
Whereas the composition profile $\overline{\phi}(z)$ (Eq.~\Eq{phiapprox}) and the electrostatic potential profile
$\overline{\psi}(z)$ (Eq.~\Eq{psiapprox}) are independent of the ion-solvent coupling $V_\pm$, the ion number density
profiles $\overline{\rho}_\pm(z)$ (Eq.~\Eq{rhoapprox}) are not.

\begin{figure}[!t]
   \includegraphics[width=8cm]{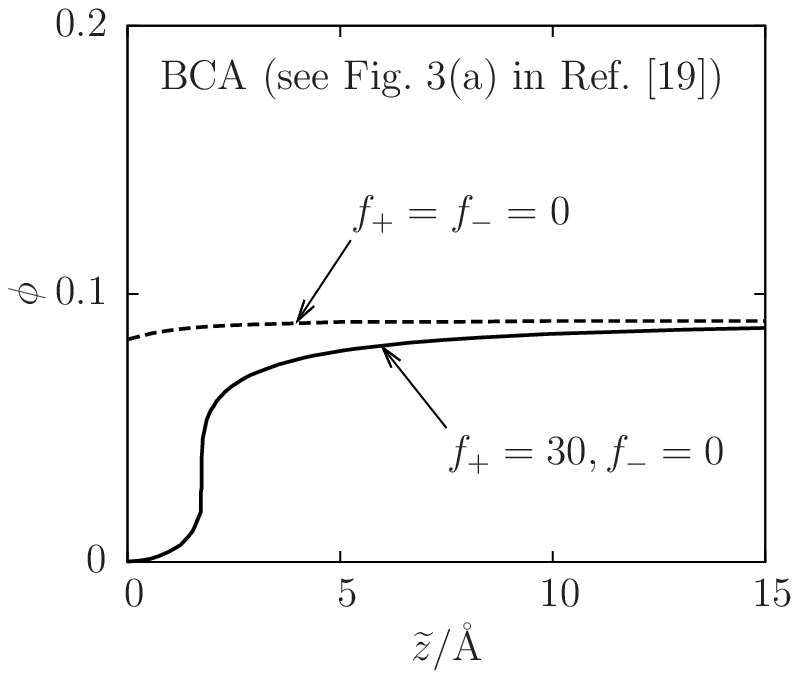}
   \caption{Composition profiles $\phi(\widetilde{z})$ within BCA for strong ion-solvent coupling ($f_+=30,f_-=0$, solid
            line) and in the absence of ion-solvent coupling ($f_+=f_-=0$, dashed line) taken from Fig.~3(a) in 
            Ref.~\cite{BenYaakov2009}.
            Here a solvent with $\epsilon_A=80$ and $\epsilon_B=20$ at bulk composition $\phi_b=0.09$ is considered.
            The bulk ionic strength is $\widetilde{I}=0.1\,\m{mM}$ and the surface charge density is
            $\widetilde{\sigma}=-16\,\m{\mu C/cm^2}$.
            At distances $\widetilde{z}<2\,\m{\AA}$ from the wall the two curves differ strongly from each other, 
            whereas the differences are small at large distances.}
   \label{fig:8}
\end{figure}

At distances from the wall of more than a few particle diameter ($z>2$) the approximate profiles 
$\overline{\phi}(z)$, $\overline{\psi}(z)$, and $\overline{\rho}_\pm(z)$ differ only slightly from the ones 
obtained by the full numerical minimization.
Closer to the wall the deviations between the numerical and the approximate profiles are more pronounced, but in 
this spatial range the present local model is not conclusive because it neglects the surface layering of actual fluids.
A similar situation occurs within BCA \cite{BenYaakov2009} (see Fig.~3(a) therein) shown in Fig.~\ref{fig:8} .
There, at distances $\widetilde{z} < 2\m{\AA}$, the solvent composition profile $\phi(\widetilde{z})$ for strong 
ion-solvent coupling ($f_+=30,f_-=0$, solid line) differs strongly from that in the absence of ion-solvent coupling 
($f_+=f_-=0$, dashed line).
At large distances the deviations are small.
A closer comparison between Fig.~\ref{fig:8} and Fig.~\ref{fig:7}(a) would require the knowledge of the particle size,
which is however not specified in Ref.~\cite{BenYaakov2009}.
A description of the presently considered semi-infinite planar system within RPA has been given in Ref.~\cite{Ciach2010}.
There the ion-solvent coupling has been treated perturbatively, but the full, numerically determined profiles $\phi(z)$, 
$\psi(z)$, and $\rho_\pm(z)$ within RPA have not been discussed.
However, by comparing these profiles, as obtained within RPA, with those obtained within BCA or LDA one could assess
the influence of non-locality on the interfacial structure in the complex fluids studied here.


\subsection{\label{Sec:CritAd}Critical adsorption} 

Here we investigate critical adsorption at a wall with a strong surface field $h$ and with surface 
charge density $\sigma$.
We consider the case that in the bulk the binary liquid mixture is at the critical bulk composition $\phi_b=\phi_c$
in the presence of salt with bulk ionic strength $\rho_{\pm b}\equiv I=I_c$ (see Subsec.~\ref{Sec:Critical}). 
A surface field $h>0$ ($h<0$) favors the adsorption of $A$ ($B$) particles and leads to a local segregation.
In order to obtain an analytical expression for the excess adsorption $\Gamma(\xi)$, which captures the full 
mean-field behavior to leading order close to the critical point ($\xi\to\infty$), we expand the density functional 
in Eq.~\Eq{df} in two steps in order to derive a Ginzburg-Landau-type description. 
In the first step the density functional $\Omega[\phi,\rho_\pm]$ in Eq.~\Eq{df} is expanded up to second order in the 
deviations $\Delta\rho_\pm(z):=\rho_\pm(z)-I$ of the ion densities from their bulk equilibrium values $I=I_c$. 
This leads to a density functional $\Omega_1[\phi,\Delta\rho_\pm]$.
Minimizing $\Omega_1[\phi,\Delta\rho_\pm]$ with respect to $\Delta\rho_\pm$ renders Euler-Lagrange 
equations linear in $\Delta\rho_\pm(z)$, the solutions of which are functionals $\Delta\rho_\pm^*(z,[\phi])$ of the 
(up to here unknown) solvent composition profile $\phi$.
Inserting the solutions $\Delta\rho_\pm^*(z,[\phi])$ into the density functional, $\Omega_1[\phi,\Delta\rho_\pm]$ and,
as the second step, expanding $\Omega_1[\phi,\Delta\rho_\pm^*[\phi]]$ up to fourth order in the order parameter deviations 
$\varphi(z):=\phi(z)-\phi_c$ leads to the Ginzburg-Landau-type functional
\begin{eqnarray}
   \frac{\mathcal{H}[\varphi]}{A} 
   & = & 
   \int_0^\infty\d z\,\Big(a\big(\varphi(z)\big)^2+b\big(\varphi(z)\big)^4 
   +c\big(\varphi'(z)\big)^2 \nonumber\\
   & &
   +U(z)\varphi(z)\Big) - h\varphi(0) + \mathcal{O}((\Delta\gamma)^2), \label{eq:gl}
\end{eqnarray} 
where $A$ is the surface area in units of $\widetilde{a}^2$.
Here $a=\chi_c-\chi$, $b=4/3$, $c=\chi/6$, and the effective ``external'' field is
\begin{equation}
   U(z) = -\frac{\kappa\sigma\Delta\gamma}{2}\exp(-\kappa z) - 
   \frac{2\pi\ell_B\sigma^2\epsilon'(\phi_c)}{(\epsilon(\phi_c))^2}\exp(-2\kappa z).
   \label{eq:U}
\end{equation}
The external field $U(z)$ describes the influence of surface charges $\sigma$ on the order parameter $\varphi$.
The first term on the right-hand side of Eq.~\Eq{U} is due to the ion solubility whereas the second term is due to the 
dielectric properties of the solvent.
Solving perturbatively to first order in $U$ the Euler-Lagrange equation, obtained from $\mathcal{H}/A$ in Eq.~\Eq{gl},
leads to the equilibrium order parameter profile $\varphi_\m{eq}(z;\xi)$.

From Eq.~\Eq{gl} one obtains the bulk correlation length 
$\xi=\lambda+\mathcal{O}((\Delta\gamma)^2)=\lambda\sqrt{1-g^2}+\mathcal{O}((\Delta\gamma)^2)=\xi^{(OZ)}+\mathcal{O}((\Delta\gamma)^2)$ with
$\dps\lambda=\sqrt{\frac{\chi}{6(\chi_c-\chi)}}=\sqrt{\frac{c}{a}}$ (see Eqs.~\Eq{lambda}, \Eq{g2}, and \Eq{xiOZ}
with $\chi_s(\phi_c,I_c)=\chi_c$) for $\chi<\chi_c$.
In the following we shall neglect the corrections $\mathcal{O}((\Delta\gamma)^2)$, which are expected to be small within LDA
(see Sec.~\ref{Sec:Bulk}).
Using the empirical form $\dps\chi(T)=\chi_S+\frac{\chi_H}{T}$ (see Sec.~\ref{Sec:Model}) one obtains 
$\xi(T\to T_c)\stackrel{t\to0}{\to}\xi_0^+|t|^{-\nu}$ with the non-universal critical amplitude $\dps\xi_0^+=\sqrt{\frac{\chi_cT_c}{6|\chi_H|}}$, the critical
(mean-field) exponent $\nu=1/2$, and $t=(T-T_c)/T_c$.
Moreover, from Eq.~\Eq{gl} one obtains the bulk order parameter 
$\dps\varphi_\m{eq}(z=\infty;\xi)=\sqrt{-\frac{a}{2b}}=\sqrt{\frac{3(\chi-\chi_c)}{8}}\stackrel{t\to0}{\to} m_0|t|^\beta$
with the critical amplitude $\dps m_0=\sqrt{\frac{3|\chi_H|}{8T_c}}$ and the critical (mean-field) exponent $\beta=1/2$.
For later purposes (see the text below Eqs.~\Eq{critadprof} and \Eq{Gamma1}) here we note that 
$\sqrt{c/b} = \sqrt{2\chi_c}/4 = \sqrt{2}m_0\xi_0^+=1/2$ for $\chi=\chi_c=2$ (see 
Sec.~\ref{Sec:Critical}).

At the critical point ($\xi=\infty$) and far away from the substrate the equilibrium order parameter profile 
$\varphi_\m{eq}(z;\xi=\infty)$ decays as
\begin{eqnarray}
   &&
   \varphi_\m{eq}(z\to\infty;\xi=\infty)
   = 
   \frac{\sign(h)}{2z}
   \label{eq:critadprof}\\
   &&
   \hspace{3em} + \Big(-\frac{\sign(h)}{2} + \frac{9\sigma\Delta\gamma}{10\kappa^3} + 
   \frac{9\pi\ell_B\sigma^2\epsilon'(\phi_c)}{40\kappa^4(\epsilon(\phi_c))^2}\Big)\frac{1}{z^2}
   \nonumber\\
   &&
   \hspace{3em} + \mathcal{O}(z^{-3}).
   \nonumber
\end{eqnarray}
The leading contribution $\dps\sign(h)\frac{\sqrt{c/b}}{z}$ can be written in the scaling form 
$\sign(h)m_0c_+(z/\xi_0^+)^{-\beta/\nu}$ with the universal amplitude $c_+=\sqrt{2}$ \cite{Floter1995}, where the 
critical exponents take their mean-field values $\beta=\nu=\frac{1}{2}$ \cite{Pelissetto2002}.
Accordingly, the leading term in Eq.~\Eq{critadprof} is not affected by the surface charge, the presence of ions, or the 
dielectric properties of the solvent.
However, these materials properties do modify the amplitude of the subleading contribution ($\propto1/z^2$).

Close to the critical point ($\xi\to\infty$) the excess adsorption $\Gamma(\xi)=\int\limits_0^\infty\!\d z\;
\varphi_\m{eq}(z;\xi)$ with the perturbatively obtained profile $\varphi_\m{eq}(z;\xi)$ (see above) is given by
\begin{equation}
   \Gamma(\xi) = \Gamma_0(\xi) + \Gamma_1(\xi) + \mathcal{O}(1/\xi) \label{eq:Gamma}
\end{equation}
with 
\begin{equation}
   \Gamma_0(\xi) := \frac{\sign(h)}{2}\ln(2\xi) 
   \label{eq:Gamma0}
\end{equation}
and
\begin{equation}
   \Gamma_1(\xi) := \frac{3\sigma\Delta\gamma}{8\kappa^2} + 
                    \frac{3\pi\ell_B\sigma^2\epsilon'(\phi_c)}{16\kappa^3(\epsilon(\phi_c))^2}.
   \label{eq:Gamma1}
\end{equation}
The leading contribution $\dps\sign(h)\sqrt{\frac{c}{b}}\ln(\xi)$ can be written in the scaling form 
$\simeq\sign(h)m_0\xi_0^+g_+(-\ln(|t|))$ with the universal amplitude $g_+=\sqrt{2}\nu=1/\sqrt{2}$ within mean-field 
theory \cite{Floter1995}.
\begin{figure}[!t]
   \includegraphics[width=8cm]{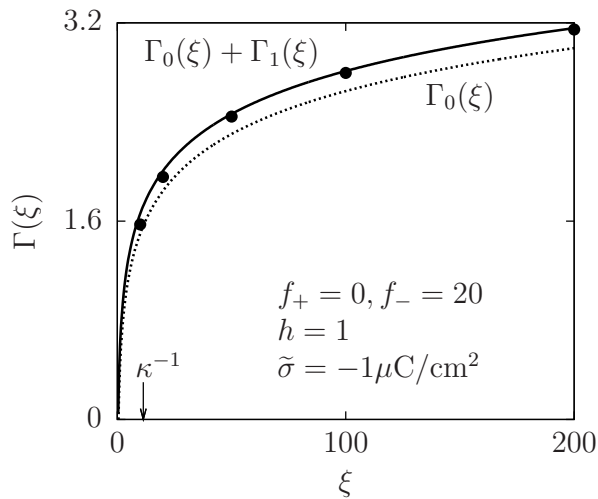}
   \caption{Comparison of the numerically calculated excess adsorption $\Gamma(\xi)$ obtained within the full model in 
            Eq.~\Eq{df} ($\bullet$) with the predictions of Eq.~\Eq{Gamma} for the parameters used in Fig.~\ref{fig:7}
            with $h=1$. 
            The Debye length $\kappa^{-1}$ (marked by an arrow) corresponds to a bulk ionic strength $\widetilde{I}=10\,\m{mM}$.
            The term $\Gamma_0(\xi)$ (dotted line, see Eq.~\Eq{Gamma0}), which contains the leading contribution to 
            the excess adsorption and which corresponds to a vanishing surface  charge density ($\sigma=0$), exhibits 
            visible deviations from the numerical results ($\bullet$); nonetheless $\Gamma_1(\xi)/\Gamma_0(\xi)\to0$ for
            $\xi\to\infty$.
            Taking into account in addition the term $\Gamma_1(\xi)$ (see Eq.~\Eq{Gamma1}), which exhibits a dependence
            on the surface charge $\sigma$, quantitative agreement is found between $\Gamma_0(\xi)+\Gamma_1(\xi)$ (solid
            line) and the numerical results ($\bullet$) in the limit $\xi\to\infty$.
            This finding also implies that those terms of Eq.~\Eq{df}, which have been left out upon deriving Eq.~\Eq{gl},
            do not contribute detectably.}
   \label{fig:9}
\end{figure}
$\Gamma_0(\xi)$ diverges for $\xi\to\infty$ whereas $\Gamma_1(\xi)$ remains finite and thus represents the first 
subdominant correction.
Figure~\ref{fig:9} compares the predictions of Eq.~\Eq{Gamma} with the results obtained by numerically calculating 
the excess adsorption within the \emph{full} model as given by Eq.~\Eq{df} ($\bullet$) for the parameters used in 
Fig.~\ref{fig:7} with $h=1$, in particular for large $\xi$.
Whereas the leading contribution $\Gamma_0(\xi)$ in Eq.~\Eq{Gamma0} itself (dotted line) deviates visibly from the 
full numerical results ($\bullet$), there is quantitative agreement between the latter and $\Gamma_0(\xi)+\Gamma_1(\xi)$
(solid line) in the limit $\xi\to\infty$.
Since $\Gamma_0(\xi)$ corresponds to a vanishing surface charge ($\sigma=0$), the difference between the dotted and the
full line in Fig.~\ref{fig:9} demonstrates the influence of electrostatic interactions on the excess adsorption.
The quantitative agreement of $\Gamma_0(\xi)+\Gamma_1(\xi)$ (solid line) with the numerical results ($\bullet$) indicates
that the terms of Eq.~\Eq{df} neglected upon deriving Eq.~\Eq{gl} do not contribute detectably to the leading and the 
first subleading behavior of $\Gamma(\xi\to\infty)$.
Moreover, for the given choice of parameters the magnitude of the correction $|\mathcal{O}(1/\xi)|$ in Eq.~\Eq{Gamma}
turns out to be smaller than $|\Gamma_1(\xi)|$, which in turn vanishes relative to $\Gamma(\xi\to\infty)$.
Note that within BCA, for $f_+=0,f_-=20$ (as used in Fig.~\ref{fig:9}) the uniform bulk state is unstable in a certain 
vicinity of the critical point, because $|g|=1.39>1$ (see Sec.~\ref{Sec:Correlations}), which precludes calculating
the excess adsorption $\Gamma(\xi)$.

Critical adsorption occurs upon approaching the critical point ($\xi\to\infty$), where the excess adsorption diverges as
$\Gamma\propto\ln\xi$, which is in agreement with the expected universal scaling behavior 
$\Gamma\propto\xi^{1-\beta/\nu}$ \cite{Dietrich1988,Floter1995} for the classical exponents $\beta=\nu=1/2$ 
corresponding to the present mean-field theory.
It is apparent from Eq.~\Eq{Gamma} that the leading contribution $\Gamma_0$ is not altered by the surface charge, the 
presence of ions, or the dielectric properties of the solvent.
However, these non-universal properties do influence the subleading contribution $\Gamma_1$.

Recently the adsorption of critical water+2,6-dimethylpyridine mixtures with $\m{KBr}$ of various ionic strengths $I$
has been investigated by means of surface plasmon resonance \cite{Nellen2011}.
For the case of a hydrophobic wall the excess adsorption turned out to be practically independent of the ionic strength.
This is in agreement with Eq.~\Eq{Gamma} because a hydrophobic wall is only weakly charged \cite{Rudhardt1998} such 
that the second and third terms on the right-hand side of Eq.~\Eq{Gamma} are negligibly small.

For the case of a hydrophilic, negatively charged ($\sigma<0$) wall a decrease of the adsorption of water has been 
measured upon adding salt \cite{Nellen2011}. 
Hydrophilic walls can be expected to be strongly charged \cite{Behrens2001} such that 
$\sigma=\sign(\sigma)\sigma_\m{sat}$ with the saturation surface charge density 
$\sigma_\m{sat}=\kappa\epsilon(\phi_c)/(\pi\ell_B)$ \cite{Bocquet2002}.
In this case from Eq.~\Eq{Gamma} one obtains
\begin{equation}
   \frac{\partial\Gamma}{\partial I} = -\frac{3}{4\kappa^3}\Big(2\sign(\sigma)\Delta\gamma + 
   \frac{\epsilon'(\phi_c)}{\epsilon(\phi_c)}\Big).
   \label{eq:GammaI}
\end{equation}
Equation~\Eq{GammaI} assumes only a weak dependence of $\phi_c$ (and thus of $\Delta\gamma$ and of $\epsilon(\phi_c)$) 
on the ionic strength $I$ (see Sec.~\ref{Sec:Critical}) so that the derivative $\dps\frac{\d\phi_c}{\d I}$ does not 
appear.

If 2,6-dimethylpyridine is denoted as the $A$ component and water as the $B$ component of the binary liquid mixture
(i.e., $\Gamma$ measures the excess of 2,6-dimethylpyridine), at the lower critical demixing point an experimental value 
of $\epsilon'/\epsilon\approx-1.2$ is found \cite{Lide1998,Kaatze1984}.
For this mixture the solubility contrasts for $\m{KBr}$ are $f_+\approx2.5$ and $f_-\approx8.4$ \cite{Inerowicz1994}
which leads to $f_+-f_-\approx-5.9$ and $\Delta\gamma\approx-0.30$.
Within LDA from these numbers one finds $\partial\Gamma/\partial I>0$ (i.e., decreasing water adsorption upon adding 
salt), and the second (dielectric) contribution on the right-hand side of Eq.~\Eq{GammaI} dominates.
In contrast, within BCA one has $\partial\Gamma^\m{(BCA)}/\partial I<0$ (i.e., increasing water adsorption upon adding
salt), because $\Delta\gamma\approx f_+-f_-$ leads to a dominance of the first (ion solubility) contribution on 
the right-hand side of Eq.~\Eq{GammaI}.
Hence the overestimation of the ion-solvent coupling within BCA leads to a sign of $\partial\Gamma^\m{(BCA)}/
\partial I$ which is not compatible with the aforementioned experimental findings in Ref.~\cite{Nellen2011}, whereas 
the sign of $\partial \Gamma/\partial I$ within the present LDA is in agreement with these findings.
Since the dielectric properties are experimentally accessible one could use Eq.~\Eq{GammaI} to determine
$\Delta\gamma$ from measurements of the excess adsorption $\Gamma$ as a function of the ionic strength $I$.
A comparison of this resulting value for $\Delta\gamma$ with the difference $f_+-f_-$ of the Gibbs free energies of 
transfer (inferred, e.g., from electrochemical methods) would be a direct way to probe quantitatively the difference 
between the LDA and the BCA.

In order to further test the different predictions following from BCA and LDA we suggest additional adsorption 
measurements for hydrophilic walls.
For $\m{KBr}$ as salt the arguments above lead to the assertions that, within LDA, one has $\partial\Gamma/\partial I>0$ 
(i.e., decreasing water adsorption upon adding salt) independently of the sign of the surface charge $\sigma$ (because the
last term on the right-hand side of Eq.~\Eq{GammaI} dominates), whereas 
within BCA $\partial\Gamma^\m{(BCA)}/\partial I$ is expected to change sign upon changing the sign of $\sigma$ (because
the first term on the right-hand side of Eq.~\Eq{GammaI} dominates).
More interestingly, using an antagonistic electrolyte (i.e., with $f_+$ and $f_-$ having opposite signs) such as 
$\m{HBr}$ ($f_+\approx-11.2,f_-\approx8.4$ \cite{Inerowicz1994}, i.e., $\Delta\gamma\approx-4$) the first (ion 
solubility) contribution on the right-hand side of Eq.~\Eq{GammaI} is dominating such that $\partial\Gamma/\partial I$
and $\sigma$ are expected to have the same sign.
In this case, upon adding salt, the amount of adsorbed water either decreases or increases depending on the sign of
the surface charge.
This is in contrast to electrowetting where the water adsorption increases with the magnitude but independent of the 
sign of the surface charge \cite{Mugele2005}.
Within the BCA approach of Ref.~\cite{Samin2011a} the difference for cations and anions with respect to their solubility 
contrasts in the two pure solvent components is neglected (i.e., $f_+=f_-\gg1$ so that $\Delta\gamma=0$) 
to the effect that the reported capillary condensation-like adsorption of water between two equally charged walls at 
variable distance should be independent of the sign of the surface charge.
The analysis above implies that the same property is expected to occur for sufficiently small values of 
$|\Delta\gamma|$, but the adsorption may depend on the sign of the surface charge if $|\Delta\gamma|$ becomes of the
order unity.


\section{\label{Sec:Colloids}Colloidal interactions in near-critical electrolyte solutions}

The interactions between colloidal particles in a fluid medium comprise dispersion forces, direct screened Coulomb
forces, steric forces, as well as solvent-mediated interactions, which are commonly referred to as solvation forces.
If the thermodynamic state of the fluid medium is moved towards a critical point, e.g., the critical demixing point of a
binary liquid mixture, the fluctuation induced long-ranged, and universal critical Casimir force emerges; this singular
contribution to the solvation force dominates \cite{Krech1994,Schlesener2003,Gambassi2009a}.
In the presence of a sufficiently strong adsorption preference of colloids or of confining walls for one of the components 
of the solvent the critical Casimir force depends only on the relative \emph{signs} of the surface fields $h$ acting 
on the order parameter at these surfaces (and on the geometry of the latter) but not on non-universal material parameters.
The critical Casimir force is attractive (repulsive) if the surface fields have equal (opposite) signs.

While being investigated theoretically for quite some time \cite{Krech1994}, the direct experimental verification of
the critical Casimir effect has been achieved only recently for a single colloidal particle dispersed in a
mixture of water and 2,6-dimethylpyridine close to a planar wall \cite{Gambassi2009a,Hertlein2008,Gambassi2009b}.
In that study the measured effective colloid-wall interaction potential was interpreted in terms of a superposition 
of a repulsive screened Coulomb force and the critical Casimir force.
Within this picture the direct electrostatic repulsion dominates for temperatures $T$ which deviate from $T_c$ more
than a few tenth of a Kelvin, whereas an increasingly strong Casimir attraction (repulsion) occurs 
for symmetric (antisymmetric) boundary conditions upon approaching the critical point ($T\to T_c$).
Recently these experiments have been modeled within RPA, which provides a satisfactory fitting of 
the experimental curves in Refs.~\cite{Hertlein2008,Gambassi2009b}.
However, this approach involves a large number of model parameters \cite{Pousaneh2012}, which limits the conclusiveness
of these fits.

The influence of adding salt onto the effective colloid-wall interaction has been studied recently \cite{Nellen2011} using
the same experimental setup as in Refs.~\cite{Hertlein2008,Gambassi2009b}.
It turns out that with $10\,\m{mM}$ of $\m{KBr}$ the Casimir attraction for symmetric boundary conditions 
starts to dominate the direct electrostatic repulsion already several Kelvin away from the critical point (instead of
tenths of a Kelvin as for the salt-free solvent).
Moreover, for antisymmetric boundary conditions, for which both the direct electrostatic and the critical Casimir 
forces are expected to be repulsive, in the presence of salt an attraction has been detected within an intermediate 
temperature range.
The latter observation demonstrates that, under certain conditions, assuming the simple superposition of the direct 
electrostatic and the critical Casimir forces is insufficient to understand the actual effective interaction.

Based on the Ginzburg-Landau-like description in Eq.~\Eq{gl}, which follows from the full model given in Eq.~\Eq{df}, we
have identified a mechanism giving rise to the aforementioned unexpected attraction for antisymmetric boundary conditions
in the presence of salt \cite{Bier2011}.
For $\Delta\gamma\not=0$ the cations and anions are separated close to the surfaces, where the order parameter 
$\varphi_\m{eq}$ is non-uniform.
This gives rise to dipolar layers, which can interact with the surface charges at the distant surface.
These dipole layers are expected to contribute significantly to the effective interaction if the direct electrostatic
interaction is weak; this is the case for antisymmetric boundary conditions, for which the hydrophobic wall is expected
to be weakly charged.
If the direct electrostatic interaction is strong, which is expected to occur for symmetric boundary conditions with 
hydrophilic surfaces, the dipolar layers do not significantly contribute to the effective interaction.
In that case the sole effect of adding salt is to reduce the Debye length and thereby to weaken the direct electrostatic
repulsion relative to the critical Casimir force.
Therefore, with salt the onset of effective attraction occurs at temperatures $T$ further away from $T_c$ 
than in the salt-free case \cite{Bier2011}.

A conceivable alternative mechanism for the emergence of an effective attraction in the case of antisymmetric boundary
conditions has been proposed which is independent of differences in the solubility of cations and anions 
\cite{Pousaneh2011}.
It has been argued within RPA that a charged wall (of either polarity) accumulates an increased number of ions compared 
to an uncharged wall.
Due to this enhanced total density of ions (which are hydrophilic independent of their sign) a charged wall should, 
from a distance, appear increasingly hydrophilic upon adding salt such that for certain system parameters an underlying 
actually hydrophobic character of a wall might be overcompensated and turn into an effectively hydrophilic wall; actually
hydrophilic walls remain so upon adding salt \cite{Pousaneh2011}.
Such a salt-induced apparent hydrophilicity, which would occur on the surfaces of all dissolved colloids, would in turn
lead to effectively symmetric boundary conditions and thus to attractive solvation forces.
However, according to Fig.~\ref{fig:9}, even in the presence of salt the excess adsorption follows the actual preference 
of the surface field, also upon approaching $T_c$.
As mentioned in Sec.~\ref{Sec:CritAd}, it is not possible to calculate the excess adsorption within BCA (and RPA) for
the parameters used in Fig.~\ref{fig:9}, because for them within this approach the uniform bulk state is thermodynamically
unstable.
Thus Fig.~\ref{fig:9} demonstrates that within LDA salt-induced apparent hydrophilicity does not occur (i.e., $\Gamma$
does not become negative).
Therefore there is reason to expect that salt-induced apparent hydrophilicity is an artifact of the BCA and the RPA.
In addition, by means of surface plasmon resonance it has been checked experimentally that the adsorption preference, 
in particularly of hydrophobic substrates, is not altered by adding salt \cite{Nellen2011}.
Therefore, for antisymmetric boundary conditions of the order parameter, there are doubts that salt-induced apparent 
hydrophilicity can serve as an explanation for the experimentally observed effective attraction within an intermediate 
temperature range.

Recently, additional numerical studies within BCA have been performed suggesting that ion-induced ''precipitation''
\cite{Okamoto2011} or non-linearities \cite{Samin2011b} influence the effective colloid-colloid interaction.
However, our results in Secs.~\ref{Sec:Bulk} and \ref{Sec:Interface} concerning the reliability of BCA point towards 
the possibility that those proposed effects are artifacts of the BCA due to an overestimation of the ion-solvent coupling.
Therefore it would be worthwhile to reconsider the aforementioned proposed effects within LDA, which has been shown to
be consistent with presently available experimental evidence.


\section{\label{Sec:Conclusions}Conclusions and Summary}

We have derived a local density approximation (LDA) for the density functional of point-like ions interacting locally
with a binary liquid mixture acting as a solvent (Fig.~\ref{fig:2}), which for large free energies of transfer of 
the ions improves the frequently used bilinear coupling approximation (BCA).
It turns out that within the proposed LDA, and in contrast to the BCA, the influence of ions on the bulk phase diagram
(Fig.~\ref{fig:1}), the critical point (Fig.~\ref{fig:3}), and the bulk structure (Figs.~\ref{fig:4}--\ref{fig:6}) is 
predicted to be weak.
This is in agreement with the presently available experimental data.
The interfacial structure at distances from a charged wall less than a few particle diameters can be predicted by neither
BCA nor LDA, because none of these two local models accounts for the layering due to packing effects which dominate
in actual fluids at such short distances.
But, within both BCA and LDA, further away from the wall the system can be described reliably in simple terms of a 
uniform permittivity and an effective surface field.
This description nonetheless captures the actual structure as obtained from the full numerical minimization 
(Fig.~\ref{fig:7}).
Upon approaching the critical point the subleading (but not the leading) contribution to critical adsorption is found 
to be sensitive to system and materials parameters such as the bulk ionic strength, solubility properties, surface 
charges, and the permittivities of the solvent components (Fig.~\ref{fig:9}).  

If a salt disturbs the molecular arrangement of the solvent molecules only at small distances, it is modeled here by 
\emph{point-like} ions which interact \emph{locally} with the solvent.
Such a salt does neither significantly alter the bulk phase behavior nor the bulk structure or the asymptotic decay 
of density profiles at walls.
But it contributes to the interfacial structures up to distances of the order of the Debye length $1/\kappa$ as well
as to critical adsorption in subleading order.
If the solvent is moved thermodynamically towards a critical point, where the bulk correlation length $\xi$ diverges, the 
ratio $1/(\kappa\xi)$ of the range $1/\kappa$ of the ion-related surface structure and $\xi$ becomes small, such that
the leading, universal critical behavior of the solvent is not altered by adding salt\cite{Bier2011}.

Whether the description of a given electrolyte solution within a local model is justified or not does not depend on the salt 
alone but on the combination of salt and solvent. 
Experimentally observed effects in binary liquid mixtures due to adding salt, such as the shift of the critical 
point, depend sensitively on the type of mixture (compare Ref.~\cite{Seah1993} for water+2,6-dimethylpyridine and 
Ref.~\cite{Sadakane2007a} for heavy water+3-methylpyridine).
Moreover, the measured critical point shifts exhibit a strong dependence on the size of the ions (compare 
Ref.~\cite{Sadakane2007a} for alkali halides and Ref.~\cite{Sadakane2007b} for sodium tetraphenylborate).
These evidences in combination with the present analysis, which implies a weak influence of the ionic \emph{charge},
lead to the conclusion that steric effects might play an important role for the ion-solvent interaction.
This interpretation is supported by reports of critical point shifts of similar magnitude in binary liquid mixtures
due to adding non-ionic impurities \cite{Hales1966}.
Consequently it appears as if it is mostly the property of an ion to be a structure maker or a structure breaker 
and only to a lesser extent its electric charge which determines the influence of a salt onto the properties of 
the solvent.
In order to obtain quantitatively reliable predictions for electrolyte solutions, a larger effort has to be devoted 
to study the steric and chemical influence of ions on the solvent.

In summary, for ions dissolved in a binary solvent the bulk phase diagrams (Fig.~\ref{fig:1}), the critical point 
shifts (Fig.~\ref{fig:3}), the bulk two-point correlation functions (Figs.~\ref{fig:4}--\ref{fig:6}), the surface
structures (Fig.~\ref{fig:7}), and critical adsorption (Fig.~\ref{fig:9}) have been investigated within a novel 
local density approximation (LDA) for the ion-solvent interaction (Fig.~\ref{fig:2}).
The commonly used bilinear coupling approximation (BCA) turns out to strongly overestimate the influence of ions on the
solvent properties in cases of realistic solvation free energies, whereas the LDA introduced here, in agreement with 
various available experimental data, predicts small effects. 
Although the presented LDA is expected to be more accurate than the BCA, the former requires the same and not more 
parameters than the latter.
According to its derivation, the LDA is expected to be reliable for small ionic strengths and on length scales larger
than the particle size.
Both available and possible future experiments have been discussed to probe and to explore the reliability and the 
range of validity of the present theoretical approach.


\begin{acknowledgments}
We thank M.\ Oettel, U.\ Nellen, J.\ Dietrich, and C.\ Bechinger for many stimulating discussions.
A.G.\ is supported by MIUR within ``Incentivazione alla mobilit\`{a} di studiosi stranieri e
italiani residenti all'estero.''
\end{acknowledgments}


\appendix

\section{\label{AppA}Derivation of the solvent-induced ion potential $V_\pm(\phi)$}

Within the present model the solvation of ions is dominated by short-ranged interactions with solvent particles.
Accordingly, the ion-solvent interaction is modeled locally in terms of a free energy density 
$\dps\sum_{i=\pm}\rho_iV_i(\phi)$ (see Eq.~\Eq{df}).
In order to derive an expression for the solvent-induced ion potential $V_\pm(\phi)$, resorting to a lattice gas model
and in line with the short range of the ion-solvent interaction, a single site is considered which is occupied by one
solvent particle (either of type $A$ or of type $B$, i.e., corresponding to occupation numbers $N_A=1-N_B\in\{0,1\}$)
and an arbitrary number of positive and negative ions ($N_\pm\in\mathbb{N}_0$).
The surrounding of this considered site acts as a particle reservoir which is described by chemical potentials 
$\lambda_ik_BT, i\in\{A,B,+,-\}$, and the interaction energy per $k_BT$ of two particles of species $i,j\in\{A,B,+,-\}$
on that site is given by $k_{ij}$ with $k_{ij}=0$ for $i,j\in\{A,B\}$, because a particle does not interact with 
itself and the site cannot be occupied by more than one particle of type $A$ or $B$. 
The chemical potentials $\lambda_i$ are related to the chemical potentials $\mu_\phi$ and $\mu_\pm$ introduced in 
Sec.~\ref{Sec:Model} (see below).
Accordingly, the Hamiltonian $Hk_BT$ of a configuration $(N_A,N_B,N_+,N_-)\in\{0,1\}\times\{0,1\}\times\mathbb{N}_0
\times\mathbb{N}_0$ on one site with $N_A+N_B=1$ is given by
\begin{eqnarray}
   H(N_A,N_B,N_+,N_-) & = & k_{A+}N_AN_+ + k_{A-}N_AN_- + \nonumber\\
   &&
   k_{B+}N_BN_+ + k_{B-}N_BN_- + \nonumber\\
   &&
   H_\m{ion}(N_+,N_-) \label{eq:H}
\end{eqnarray}
with the ionic part
\begin{eqnarray}
   &&
   H_\m{ion}(N_+,N_-) = \label{eq:Hion}\\
   &&
   k_{++}\frac{N_+(N_+-1)}{2} + k_{+-}N_+N_- + k_{--}\frac{N_-(N_--1)}{2}. \nonumber
\end{eqnarray}
The corresponding grand partition function is
\begin{eqnarray}
   && 
   \zeta(\lambda_A,\lambda_B,\lambda_+,\lambda_-) = \sum_{(N_A,N_B,N_+,N_-)} \label{eq:partsum1}\\
   &&
   \phantom{M}\exp\Big(\sum_{i\in\{A,B,+,-\}} \lambda_iN_i - H(N_A,N_B,N_+,N_-)\Big),\nonumber   
\end{eqnarray} 
where the outermost summation is over all configurations $(N_A,N_B,N_+,N_-)\in\{0,1\}\times\{0,1\}\times\mathbb{N}_0
\times\mathbb{N}_0$ with $N_A+N_B=1$.
With $N_A=N_\phi, N_B=1-N_\phi$ for $N_\phi\in\{0,1\}$ the Hamiltonian in Eq.~\Eq{H} can be rewritten as
\begin{eqnarray}
   H(N_\phi,1-N_\phi,N_+,N_-) & = & f_+N_\phi N_+ + f_-N_\phi N_- + \nonumber\\
   &&
   k_{B+}N_+ + k_{B-}N_- + \nonumber\\
   &&
   H_\m{ion}(N_+,N_-)
\end{eqnarray}
with the solvation energy difference $f_\pm = k_{A\pm} - k_{B\pm}$ of a $\pm$-ion.
The grand partition function in Eq.~\Eq{partsum1} is 
$\zeta(\lambda_A,\lambda_B,\lambda_+,\lambda_-)=\exp(\lambda_B)Z(\mu_\phi,\mu_+,\mu_-)$ with
\begin{eqnarray}
   && 
   Z(\mu_\phi,\mu_+,\mu_-) = \sum_{(N_\phi,N_+,N_-)}\exp\Big(\sum_{i\in\{\phi,+,-\}} \mu_iN_i \nonumber\\
   &&
   \phantom{M} - f_+N_\phi N_+ - f_-N_\phi N_- - H_\m{ion}(N_+,N_-)\Big), \label{eq:partsum2}
\end{eqnarray} 
where
$\mu_\phi = \lambda_A-\lambda_B$ and $\mu_\pm = \lambda_\pm - k_{B\pm}$ (see Sec.~\ref{Sec:Model})
and where the outermost summation is over all configurations 
$(N_\phi,N_+,N_-)\in\{0,1\}\times\mathbb{N}_0\times\mathbb{N}_0$.

In the limit of low ionic strength ($\mu_\pm\to-\infty$; this implies $\lambda_\pm\to-\infty$ which in turn 
means that the averages $\langle N_\pm \rangle=\rho_\pm$ are small) one obtains
\begin{eqnarray}
   Z(\mu_\phi,\mu_+,\mu_-)
   & \simeq & 1 + \exp(\mu_\phi) + \\
   &  & 
   \exp(\mu_+)(1 + \exp(\mu_\phi-f_+)) + \nonumber\\
   & &
   \exp(\mu_-)(1 + \exp(\mu_\phi-f_-)). \nonumber
\end{eqnarray}
Using $\dps\phi=\langle N_\phi \rangle=\frac{\partial\ln Z}{\partial \mu_\phi}$ and $\dps\rho_\pm=\langle N_\pm 
\rangle=\frac{\partial\ln Z}{\partial \mu_\pm}$ one finds from the grand canonical potential 
$-k_BT\ln Z(\mu_\phi,\mu_+,\mu_-)$ the Helmholtz free energy per $k_BT$
\begin{eqnarray}
   & & 
   -\ln Z(\mu_\phi,\mu_+,\mu_-) + \phi \mu_\phi + \rho_+\mu_+ + \rho_-\mu_- \nonumber\\
   & \simeq &
   \phi\ln\phi + (1-\phi)\ln(1-\phi) + \label{eq:sitefreeenergy}\\
   & &
   \rho_+(\ln\rho_+ - 1 + \ln M_+) + \rho_-(\ln\rho_- - 1 + \ln M_+) \nonumber
\end{eqnarray}
with $M_\pm = (1-\phi(1-\exp(-f_\pm)))^{-1}$.
From Eq.~\Eq{sitefreeenergy} one infers the effective ion-solvent interaction $V_\pm(\phi):=\ln M_\pm$.



\end{document}